\documentclass[psfig, graphics, aps]{article}
\usepackage{graphics}
\usepackage{setspace}

\makeatletter

\providecommand{\LyX}{L\kern-.1667em\lower.25em\hbox{Y}\kern-.125emX\@}

\newcommand{\lyxaddress}[1]{
  \par {\raggedright #1 
  \vspace{1.4em}
  \noindent\par}
}

\makeatother

\begin{document}

\title{Fractal Dimension of Higher-Dimensional Chaotic Repellors}

\author{David Sweet and Edward Ott\thanks{
Also Department of Electrical Engineering
}}

\maketitle

\lyxaddress{Institute for Plasma Research and Department of Physics \\
University of Maryland, College Park, Maryland 20742\\
dsweet@chaos.umd.edu\\
Ph: (301)405-1657~~~~~Fax: (301)405-1678}

\begin{abstract}
Using examples we test formulae previously conjectured to give the fractal information
dimension of chaotic repellors and their stable and unstable manifolds in ``typical''
dynamical systems in terms of the Lyapunov exponents and the characteristic
escape time from the repellor. Our main example, a three-dimensional chaotic
scattering billiard, yields a new structure for its invariant manifolds. This
system also provides an example of a system which is not typical and illustrates
how perturbation to the system restores typicality and the applicability of
the dimension formulae.
\end{abstract}

\section{Introduction}

In addition to chaotic attractors, nonattracting chaotic sets (also called chaotic
saddles or chaotic repellors) are also of great practical importance. In particular,
such sets arise in the consideration of chaotic scattering, boundaries between
basins of attraction, and chaotic transients. If a cloud of initial conditions
is sprinkled in a bounded region including a nonattracting chaotic set, the
orbits originating at these points eventually leave the vicinity of the set,
and there is a characteristic escape time, \( \tau  \), such that, at late
time, the fraction of the cloud still in the region decays exponentially at
the rate \( \tau ^{-1} \).

The primary focus of this paper will be on studying the fractal dimension of
nonattracting chaotic sets and their stable and unstable manifolds. Fractal
dimension is of basic interest as a means of characterizing the geometric complexity
of chaotic sets. In addition, a knowledge of the fractal dimension can, in some
situations, provide quantitative information that is of potential practical
use. For example, in the case of boundaries between different basins, the basin
boundary is typically the stable manifold of a nonattracting chaotic set, and
knowledge of the stable manifold's box-counting dimension (also called its capacity)
quantifies the degree to which uncertainties in initial conditions result in
errors in predicting the type of long-term motion that results (e.g., which
attractor is approached; see Sec. 5 and Ref. \cite{McDGY}). Our focus is on
obtaining the information dimension of a suitable ``natural measure'' \( \mu  \)
lying on the chaotic set (see Sec. 2 for definitions of the natural measures
for nonattracting chaotic sets and their stable and unstable manifolds). The
information dimension is a member of a one parameter (\( q \)) class of dimension
definitions given by
\begin{equation}
\label{DsubQ}
D_{q}=\lim _{\epsilon \rightarrow 0}\frac{1}{1-q}\frac{\ln \sum \mu ^{q}_{i}}{\ln (1/\epsilon )},
\end{equation}
where \( q \) is a real index, \( \epsilon  \) is the grid spacing for a \( d \)-dimensional
rectangular grid dividing the \( d \)-dimensional state space of the system,
and \( \mu _{i} \) is the natural measure of the \( i^{\textrm{th}} \) grid
cube. The box-counting dimension is given by (\ref{DsubQ}) with \( q=0 \),
and the information dimension is given by taking the limit \( q\rightarrow 1 \)
in (\ref{DsubQ}),
\begin{equation}
\label{Dsub1}
D_{1}=\lim _{\epsilon \rightarrow 0}\frac{I(\epsilon )}{\ln (1/\epsilon )},\, I(\epsilon )=\sum _{i}\mu _{i}\ln (1/\mu _{i}).
\end{equation}
In general, the information dimension is a lower bound on the box-counting dimension,
\( D_{1}\leq D_{0} \). In practice, in cases where \( D_{1} \) and \( D_{0} \)
have been determined for chaotic sets, it is often found that their values are
very close.

We are specifically concerned with investigating formulae conjectured in Ref.
\cite{HOY} that give the information dimensions for the nonattracting chaotic
set and its stable and unstable manifolds in terms of Lyapunov exponents and
the decay time, \( \tau  \). These formulae generalize previous results for
nonattracting chaotic sets of two-dimensional maps with one positive and one
negative Lyapunov exponent \cite{KG,HOG}, and for Hamiltonian systems of arbitrary
dimensionality \cite{DingOtt}. In turn, these past results for nonattracting
chaotic sets were motivated by the Kaplan-Yorke conjecture which gives the information
dimension of a chaotic attractor in terms of its Lyapunov exponents \cite{KY}.
A rigorous result for the information dimension of an ergodic invariant chaotic
set of a two-dimensional diffeomorphism has been given by L.-S. Young \cite{Young},
and this result supports the Kaplan-Yorke conjecture for attractors and the
two-dimensional map results of Refs. \cite{KG} and \cite{HOG} for nonattracting
chaotic sets. Furthermore, Ledrappier \cite{L} has proven that the conjectured
Kaplan-Yorke dimension formula is an upper bound on the information dimension
of chaotic attractors. Soluble examples for typical systems and numerical experiments
tend to support the Kaplan-Yorke result. On the other hand, some specific examples
violating the Kaplan-Yorke formula also exist. Thus, in the Kaplan-Yorke conjecture
they claim that their formula for the dimension applies for ``typical'' systems.
That is, if a specific system violates the formula then by making a ``typical''
arbitrarily small change to the system we create a new system for which the
Kaplan-Yorke formula holds (this is discussed in Refs. \cite{AM-PY} and \cite{KM-PY}).
Specific systems violating the conjecture are called \emph{atypical}, and, by
Ledrappier's result, have dimensions below the Kaplan-Yorke prediction. The
conjectured results of Ref. \cite{HOY} for nonattracting chaotic sets are also
of this character in that they are also claimed to hold only for ``typical''
systems. Thus, the issue of what constitutes a typical system (or a typical
perturbation of an atypical system) is central. At present there is no rigorous
formulation of ``typical'' for this purpose. For this reason it is important
to address this question through examples. One of the purposes of this paper
is to do just that. More generally, although many examples, both analytical
and numerical, exist which support the dimension formula for chaotic attractors,
for the case of nonattracting chaotic sets there are virtually no supporting
examples (except for the case of two-dimensional maps with one positive and
one negative Lyapunov exponent \cite{KG,HOG}). This paper provides such examples.

In Section 2 we review the results of Ref. \cite{HOY}. We define the \emph{natural
transient measure} on nonattracting chaotic sets and on their stable and unstable
manifolds, and give the conjectured information dimension formulae for these
measures.

In Section 3 we introduce a noninvertible two-dimensional map with two expanding
(positive) Lyapunov exponents. This map is simple enough that it can be fully
analyzed. This analysis is especially useful in elucidating certain aspects
of the general problem, particularly in relation to the natural transient measure
and the role of fluctuation in the finite time Lyapunov exponents.

In Section 4 we introduce our main example, a hard walled billiard chaotic scattering
system in three spatial dimensions. For a particular symmetric configuration
of the scatterer, we find (Secs. 5 and 6) that the situation is atypical (i.e.,
the conjectured formula does not apply), but that deviation of the configuration
from the symmetric case immediately restores the validity of the conjectured
dimension formula.

Section 5 presents numerical results for the dimension in the atypical (symmetric)
and typical (asymmetric) cases obtaining good agreement with theoretical results
in both cases.

Section 6 gives a theory for the structure of the nonattracting chaotic invariant
set present in examples modeling our billiard scatterer. It is shown that the
stable manifold of the invariant set is a continuous, nowhere-differentiable
surface in both the typical and the atypical cases. Furthermore, we are able
to obtain an explicit formula for the dimension in the atypical case (this formula
is compared to numerical computations in Sec. 5).

In Sec. 7.1 we present a derivation of the dimension formulae of Sec. 2. Although
these formulae have been previously derived in Ref. \cite{HOY}, the previous
derivation assumed a specific type of mapping. The situation envisioned in Ref.
\cite{HOY} does not apply to our billiard example. Thus, we are motivated to
provide more general arguments for the applicability of the formulae in Sec.
2. In Sec. 7.2 we derive a formula for the dimension of the stable manifold
of the chaotic repellor found in the typical (asymmetric) case. This derivation
shows explicitly where the derivation of the typical formula fails for this
nontrival atypical system.

In Sec. 8 we present conclusions and further discussion. In particular, we point
out that the continuous, nowhere-differentiable surface forming the stable manifold
of our billiard example may be amenable to experimental observation in a situation
where the billiard walls are mirrors and the orbits are light rays. In general,
we comment that we believe that such optical billiard experiments provide a
particularly convenient arena for the experimental investigation of different
types of fractal basin boundaries. This has so far proven to be very difficult
in other experimental settings.

\section{Dimension Formulae}

A \emph{chaotic} \emph{saddle}, \( \Lambda  \), is a \emph{nonattracting},
ergodic, \emph{invariant} set. By \emph{invariant} we mean that all forward
and reverse time evolutions of points in \( \Lambda  \) are also in \( \Lambda  \).
The \emph{stable} \emph{manifold} of \( \Lambda  \) is the set of all initial
\emph{}conditions \emph{}which \emph{}converge to \emph{\( \Lambda  \)} upon
\emph{}forward time evolution. The \emph{unstable} \emph{manifold} of \( \Lambda  \)
is the set of all initial conditions which converge to \( \Lambda  \) upon
reverse time evolution. We say \( \Lambda  \) is \emph{nonattracting} if it
does not completely contain its unstable manifold. In such a case there are
points not in \( \Lambda  \) that converge to it on backwards iteration. 

To define the characteristic escape time, \( \tau  \), first define a bounded
region, \( R \), which contains \( \Lambda  \) and no other chaotic saddle.
Uniformly sprinkle a large number, \( N(0) \), of initial conditions in \( R \).
(In this section we take the dynamical system to be a discrete time system,
i.e., a map.) Iterate the sprinkled initial conditions forward \( n\gg 1 \)
times and discard all orbits which are no longer in \( R \). Denote the remaining
number of orbits \( N(n) \). We define \( \tau  \) as
\begin{equation}
\label{tauDecay}
e^{-n/\tau }\sim \frac{N(n)}{N(0)},
\end{equation}
or, more formally, \( \tau =\lim _{n\rightarrow \infty }\lim _{N(0)\rightarrow \infty }\ln [N(0)/N(n)]/n \).
The Lyapunov exponents are defined with respect to the \emph{natural transient
measure} of the chaotic saddle \cite{OttBook}. This measure is defined on an
open set \( C\subset R \) as
\begin{equation}
\label{MeasureDef}
\mu (C)=\lim _{n\rightarrow \infty }\lim _{N(0)\rightarrow \infty }\frac{N(\xi n,n,C)}{N(n)},
\end{equation}
where \( 0<\xi <1 \), and \( N(m,n,C) \) is the number of sprinkled orbits
still in \( R \) at time \( n \) that are also in \( C \) at the earlier
time \( m<n \). The above definition of \( \mu (C) \) is presumed to be independent
of the choice of \( \xi  \) as long as \( 0<\xi <1 \) (e.g., \( \xi =1/2 \)
will do).

We take the system to be \( M \)-dimensional with \( U \) positive and \( S \)
negative Lyapunov exponents measured with respect to \( \mu  \) (where \( U+S=M \))
which we label according to the convention,
\[
h^{+}_{U}\geq h^{+}_{U-1}\geq \cdot \cdot \cdot \geq h^{+}_{1}>0>-h^{-}_{1}\geq \cdot \cdot \cdot \geq -h^{-}_{S-1}\geq -h^{-}_{S}.\]
Following \cite{HOY} we define a forward entropy,
\[
H=\sum ^{U}_{i=1}h^{+}_{i}-\tau ^{-1}.\]

We now define a natural transient measure \( \mu _{S} \) on the stable manifold
and a natural transient measure \( \mu _{U} \) on the unstable manifold. Using
the notation of Eq. (\ref{MeasureDef}),
\begin{equation}
\label{muSubS}
\mu _{S}(C)=\lim _{n\rightarrow \infty }\lim _{N(0)\rightarrow \infty }\frac{N(0,n,C)}{N(n)},
\end{equation}
\begin{equation}
\label{muSubU}
\mu _{U}(C)=\lim _{n\rightarrow \infty }\lim _{N(0)\rightarrow \infty }\frac{N(n,n,C)}{N(n)}.
\end{equation}
Thus, considering the \( N(n) \) orbits that remain in \( R \) up to time
\( n \), the fraction of those orbits that initially started in \( C \) gives
\( \mu _{S}(C) \), and the fraction that end up in \( C \) at the final time
\( n \) gives \( \mu _{U}(C) \). We use the measure (\ref{MeasureDef}), (\ref{muSubS}),
(\ref{muSubU}) to define the information dimensions of the invariant set, the
stable manifold, and the unstable manifold, respectively.

According to Ref. \cite{HOY}, the dimension of the unstable manifold is then
\begin{equation}
\label{Dimension_Unstable}
D_{U}=U+I+\frac{H-(h^{-}_{1}+\cdot \cdot \cdot +h^{-}_{I})}{h^{-}_{I+1}},
\end{equation}
where \( I \) is defined by
\[
h^{-}_{1}+\cdot \cdot \cdot +h^{-}_{I}+h^{-}_{I+1}\geq H\geq h^{-}_{1}+\cdot \cdot \cdot +h^{-}_{I}.\]
The dimension of the stable manifold is \cite{HOY}
\begin{equation}
\label{Dimension_Stable}
D_{S}=S+J+\frac{H-(h^{+}_{1}+\cdot \cdot \cdot +h^{+}_{J})}{h^{+}_{J+1}},
\end{equation}
where \( J \) is defined by
\[
h^{+}_{1}+\cdot \cdot \cdot +h^{+}_{J}+h^{+}_{J+1}\geq H\geq h^{+}_{1}+\cdot \cdot \cdot +h^{+}_{J}.\]
Considering the chaotic saddle to be the (generic) intersection of its stable
and unstable manifolds, the generic intersection formula gives the dimension
of the saddle,
\begin{equation}
\label{Dimension_Saddle}
D_{\Lambda }=D_{U}+D_{S}-M.
\end{equation}

It is of interest to discuss some special cases of Eqs. (\ref{Dimension_Unstable})--(\ref{Dimension_Saddle}).
In the case of a chaotic attractor, the invariant set is the attractor itself,
the stable manifold is the basin of attraction, and we identify the unstable
manifold with the attractor. Thus \( D_{S}=M \) and \( D_{\Lambda }=D_{U} \).
Since points near the attractor never leave, we have \( \tau =\infty  \). Equation
(\ref{Dimension_Unstable}) then yields the Kaplan-Yorke formula \cite{KY},

\begin{equation}
\label{Dimension(KYFormula)}
D_{\Lambda }=U+I+\frac{(h^{+}_{1}+\cdots +h^{+}_{U})-(h^{-}_{1}+\ldots +h^{-}_{I})}{h^{-}_{I+1}},
\end{equation}
where \( I \) is the largest integer for which \( (h^{+}_{1}+\ldots +h^{+}_{U})-(h^{-}_{1}+\ldots +h^{-}_{I}) \)
is positive.

In the case of a two-dimensional map with one positive Lyapunov exponent \( h^{+}_{1} \)
and one negative Lyapunov exponent \( h_{1}^{-} \) with the exponents satisfying
\( h^{+}_{1}-h^{-}_{1}-1/\tau \leq 0 \), Eqs. (\ref{Dimension_Unstable}) and
(\ref{Dimension_Stable}) give the result of Ref. \cite{KG} and \cite{HOG},
\[
D_{U}=1+\frac{h^{+}_{1}-1/\tau }{h^{-}_{1}},\]
\[
D_{S}=1+\frac{h^{+}_{1}-1/\tau }{h^{+}_{1}}.\]
Another case is that of a nonattracting chaotic invariant set of a one-dimensional
map. In this case \( S=0 \) and \( U=1 \). The unstable manifold of the invariant
set has dimension one, \( D_{U}=1 \). Recalling the definition of the stable
manifold as the set of points that approach the invariant set as time increases,
we can identify the stable manifold with the invariant set itself. This is because
points in the neighborhood of the invariant set are repelled by it unless they
lie precisely on the invariant set. Thus, \( D_{S}=D_{\Lambda } \), and from
(\ref{Dimension_Stable}) and (\ref{Dimension_Saddle}) we have \( D_{S}=D_{\Lambda }=H/h_{1}^{+} \),
where \( H=h^{+}_{1}-1/\tau  \).

Still another simple situation is the case of a two-dimensional map with two
positive Lyapunov exponents. This case is particularly interesting because we
will be able to use it (Sec. 3) to gain understanding of the nature of the natural
measure whose dimension we are calculating. In this case \( U=2 \) and \( S=0 \).
Thus \( D_{U}=2 \) and \( D_{S}=D_{\Lambda } \). There are two cases {[}corresponding
to \( J=0 \) and \( J=1 \) in Eq. (\ref{Dimension_Unstable}){]}. For \( h^{+}_{2}\tau \leq 1 \),
we have that \( D_{S}=D_{\Lambda } \) is between zero and one, 
\begin{equation}
\label{Dimension_SM2Da}
D_{S}=D_{\Lambda }=1+\frac{h^{+}_{2}}{h^{+}_{1}}-\frac{1}{h^{+}_{1}\tau }.
\end{equation}
For \( h^{+}_{2}\tau \geq 1 \), we have that \( D_{S}=D_{\Lambda } \) is between
one and two,
\begin{equation}
\label{Dimension_SM2Db}
D_{S}=D_{\Lambda }=2-\frac{1}{h_{2}^{+}\tau }.
\end{equation}
In the next section we will be concerned with testing and illustrating Eqs.
(\ref{Dimension_SM2Da}) and (\ref{Dimension_SM2Db}) by use of a simple model.

\section{Illustrative Expanding Two-Dimensional Map Model}

We consider the following example,
\begin{equation}
\label{2DMap-1}
x_{n+1}=2x_{n}\, \textrm{modulo}\, 1,
\end{equation}
\begin{equation}
\label{2DMap-2}
y_{n+1}=\lambda (x_{n})y_{n}+\frac{\eta }{2\pi }\sin (2\pi x_{n}),
\end{equation}
where \( \lambda (x)>1 \), and the map is defined on the cylinder \( -\infty \leq y\leq +\infty  \),
\( 1\geq x\geq 0 \), with \( x \) regarded as angle-like. We take \( \lambda (x) \)
to be the piecewise constant function,
\begin{equation}
\label{2DMapa}
\lambda (x)=\left\{ \begin{array}{ll}
\lambda _{1} & 0<x<1/2,\\
\lambda _{2} & 1/2<x<1,
\end{array}\right. 
\end{equation}
and, without loss of generality, we assume \( \lambda _{1}\leq \lambda _{2} \).
{[}Later, in Sec. 3.5, we will consider the problem with a general function
\( \lambda (x) \) and a general chaotic map \( x_{n+1}=M(x_{n}) \) replacing
(\ref{2DMap-1}), but for now we focus on (\ref{2DMap-1})--(\ref{2DMapa}).{]}

\begin{figure}
{\par\centering \resizebox*{0.9\columnwidth}{!}{\includegraphics{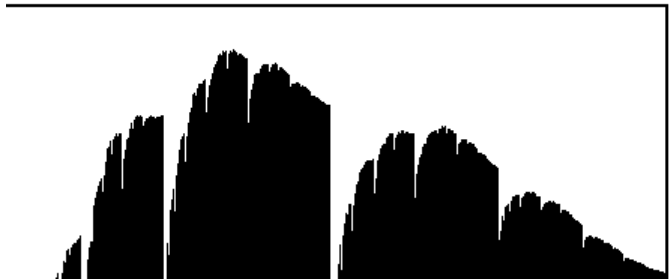}} \par}

\caption{\label{fig: 2D Fractal}Basins of 2D map model, Eqs. (\ref{2DMap-1})-(\ref{2DMapa})
for \protect\( \lambda _{1}=1.1\protect \), \protect\( r=3\protect \). A 500x500
grid of initial conditions was chosen in \protect\( 0\leq x\leq 1\protect \),
\protect\( -0.3\leq y\leq .2\protect \). Each was iterated forward by the 2D
map until the orbit fell in one of the regions \protect\( \left| y\right| >5\protect \).
If the orbit fell in \protect\( y>5\protect \) (\protect\( y<-5\protect \))
a white (black) point was plotted on the initial condition grid shown above.
The dimension of the boundary between black and white (the basin boundary) is
\protect\( D\approx 1.28\protect \).}
\end{figure}

For this map, almost every initial condition generates an orbit that either
tends toward \( y=+\infty  \) or toward \( y=-\infty  \). Figure \ref{fig: 2D Fractal}
shows the regions where initial conditions generate these two outcomes, with
the black (white) region corresponding to orbits that tend toward \( y=-\infty  \)
(\( y=+\infty  \)). Initial conditions on the border of these two regions stay
on the border forever. Thus, the border is an invariant set. It is also ergodic
by virtue of the ergodicity of the map \( x_{n+1}=2x_{n}\, mod\, 1 \). We wish
to apply Eqs. (\ref{Dimension_SM2Da}) and (\ref{Dimension_SM2Db}) to this
invariant set and its natural measure.

The Jacobian matrix for our model is
\[
{\mathcal{J}}(x)=\left[ \begin{array}{ll}
2 & 0\\
\eta \cos 2\pi x & \lambda (x)
\end{array}\right] .\]
Thus, for an ergodic invariant measure of the map, the two Lyapunov exponents
are
\begin{equation}
\label{ha}
h_{a}=p\ln \lambda _{1}+(1-p)\ln \lambda _{2}
\end{equation}
and
\[
h_{b}=\ln 2,\]
where \( p \) is the measure of the region \( x<1/2 \). To find \( h_{a} \)
we thus need to know the measure of the invariant set. The measure we are concerned
with is the natural transient measure introduced in Sec. 2.

\subsection{The Decay Time and the Natural Measure}

Consider a vertical line segment of length \( \ell _{0} \) whose \( x \) coordinate
is \( x_{0} \) and whose center is at \( y=y_{0} \). After one iterate of
the map (\ref{2DMap-1})--(\ref{2DMapa}), this line segment will have length
\( \ell _{1}=\lambda (x_{0})\ell _{0} \) and be located at \( x=x_{1} \) with
its center at \( y=y_{1} \), where \( (x_{1},y_{1}) \) are the iterates of
\( (x_{0},y_{0}) \) using the map (\ref{2DMap-1})--(\ref{2DMapa}). Thus we
see that vertical line segments are expanded by the multiplicative factor \( \lambda (x)\geq \lambda _{1}>1 \).
Now consider the strip, \( -K\leq y\leq K \), and sprinkle many initial conditions
uniformly in this region with density \( \rho _{0} \). A vertical line segment,
\( x=x_{0} \), \( -K\leq y\leq K \), iterates to \( x=x_{1} \) and with its
center at \( y_{1}=(\eta /2\pi )\sin 2\pi x_{0} \). We choose \( K>(\eta /2\pi )(\lambda _{1}-1)^{-1} \)
so that the iterated line segment spans the strip \( -K\leq y\leq K \). After
one iterate, the density will still be uniform in the strip: The region \( x<1/2 \)
\( (x>1/2) \), \( -K\leq y\leq K \), is expanded uniformly vertically by \( \lambda _{1} \)
\( (\lambda _{2}) \) and horizontally by \( 2 \). Thus, after one iterate,
the new density in the strip is \( \rho _{1}=[(\lambda _{1}^{-1}+\lambda _{2}^{-1})/2]\rho _{0} \),
and, after \( n \) iterates, we have
\[
\rho _{n}=[(\lambda _{1}^{-1}+\lambda _{2}^{-1})/2]^{n}\rho _{0}.\]
Hence the exponential decay time for the number of orbits remaining in the strip
is
\begin{equation}
\label{tauinv}
\frac{1}{\tau }=\ln \left[ \frac{1}{2}\left( \frac{1}{\lambda _{1}}+\frac{1}{\lambda _{2}}\right) \right] ^{-1}.
\end{equation}

To find the natural stable manifold transient measure of any \( x \)-interval
\( s^{(n)}_{m}=[m/2^{n},(m+1)/2^{n}] \), where \( m=0,1,\ldots ,2^{n}-1 \),
we ask what fraction of the orbits that were originally sprinkled in the strip
and are still in the strip at time \( n \) started in this interval. Let \( s^{(n)}_{m} \)
experience \( n_{1}(m) \) vertical stretches by \( \lambda _{1} \) and \( n_{2}(m)=n-n_{1}(m) \)
vertical stretches by \( \lambda _{2} \). Then the initial subregion of the
\( s^{(n)}_{m} \) still in the strip after \( n \) iterates has vertical height
\( K\lambda _{1}^{-n_{1}(m)}\lambda _{2}^{-n_{2}(m)} \). Hence the natural
measure of \( s_{m}^{(n)} \) is 
\begin{equation}
\label{musm2D}
\mu (s^{(n)}_{m})=\frac{2^{-n}\lambda _{1}^{-n_{1}(m)}\lambda _{2}^{-n_{2}(m)}}{\left[ \frac{1}{2}(\lambda _{1}^{-1}+\lambda _{2}^{-1})\right] ^{n}}=\frac{\lambda _{1}^{n_{2}(m)}\lambda _{2}^{n_{1}(m)}}{(\lambda _{1}+\lambda _{2})^{n}}.
\end{equation}
(Note that this is consistent with \( \mu ([0,1])=\sum ^{n-1}_{m=0}\mu (s_{m}^{(n)})=1 \).)
Thus the measures of the intervals \( [0,1/2] \) and \( [1/2,1] \) are
\[
p=\mu (s^{(1)}_{0})=\frac{\lambda _{2}}{\lambda _{1}+\lambda _{2}}\]
and
\[
1-p=\mu (s^{(1)}_{1})=\frac{\lambda _{1}}{\lambda _{1}+\lambda _{2}}.\]

It is important to note that our natural transient measures \( p \) and \( (1-p) \)
for the two-dimensional map are different from the natural measures of the same
\( x \)-intervals for the one-dimensional map, \( x_{n+1}=2x_{n}\, mod\, 1 \),
alone. In that case, with probability one, a random choice of \( x_{0} \) produces
an orbit which spends half its time in \( [0,1/2] \) and half its time in \( [1/2,1] \),
so that in this case the natural measures of these regions are \( p=(1-p)=1/2 \).
The addition of the \( y \)-dynamics changes the natural measure of \( x \)-intervals.

From (\ref{ha}) we obtain
\begin{equation}
\label{ha2}
h_{a}=\frac{\lambda _{2}}{\lambda _{1}+\lambda _{2}}\ln \lambda _{1}+\frac{\lambda _{1}}{\lambda _{1}+\lambda _{2}}\ln \lambda _{2}.
\end{equation}
For a general function \( f(Z) \) with \( d^{2}f/dZ^{2}<0 \), averaging over
different values of \( Z \) gives the well-known inequality \( \langle f(Z)\rangle \leq f(\langle Z\rangle ) \)
where \( \langle (\cdots )\rangle  \) denotes the average of the quantity \( (\cdots ) \).
Using \( f(Z)=\ln Z \) with \( Z=\lambda _{1} \) with probability \( p=\lambda _{2}/(\lambda _{1}+\lambda _{2}) \)
and \( Z=\lambda _{2} \) with probability \( (1-p) \), this inequality and
Eqs. (\ref{tauinv}) and (\ref{ha2}) yield the result that 
\begin{equation}
\label{ha_lt_tauinv}
h_{a}\leq \frac{1}{\tau }.
\end{equation}
In fact, we will see in Sec. 3.5 that (\ref{ha_lt_tauinv}) remains true for
any choice of the function \( \lambda (x)>1 \) in (\ref{2DMap-2}). In (\ref{ha_lt_tauinv})
the equality sign applies if the vertical stretching is uniform (\( \lambda _{1}=\lambda _{2} \))
but, for any nonuniformity in the vertical stretching (\( \lambda _{1}\neq \lambda _{2} \)),
\( h_{a} \) is strictly less than \( 1/\tau  \).

\subsection{Application of the Dimension Formulae}

Let \( \lambda _{2}=r\lambda _{1} \), \( r>1 \), and imagine that we fix \( r \)
and vary \( \lambda _{1} \). Applying Eqs. (\ref{Dimension_SM2Da}) and (\ref{Dimension_SM2Db})
to our example we obtain three cases,

(a) \( h_{b}>1/\tau >h_{a} \) (\( \lambda _{1} \) small), 

(b) \( 1/\tau >h_{b}>h_{a} \) (\( \lambda _{1} \) moderate), and

(c) \( 1/\tau >h_{a}>h_{b} \) (\( \lambda _{1} \) large).\\
Corresponding to these three cases (\ref{Dimension_SM2Da}) and (\ref{Dimension_SM2Db})
yield the following values for \( D_{\Lambda } \), the dimension of the invariant
set, 
\begin{equation}
\label{Da}
D_{a}=1+\frac{\ln (1+r^{-1})-\ln \lambda _{1}}{\ln 2},\, \textrm{for}\, \lambda _{1}\leq \lambda _{a},
\end{equation}
 
\begin{equation}
\label{Db}
D_{b}=\frac{\ln (1+r^{-1})+(1+r)^{-1}\ln r}{\ln \lambda _{1}+(1+r)^{-1}\ln r},\, \textrm{for}\, \lambda _{a}\leq \lambda _{1}\leq \lambda _{b},
\end{equation}
\begin{equation}
\label{Dc}
D_{c}=\frac{\ln (1+r^{-1})+(1+r)^{-1}\ln r}{\ln 2},\, \textrm{for}\, \lambda _{b}\leq \lambda _{1},
\end{equation}
where \( \ln \lambda _{a}=\ln (1+r^{-1}) \) and \( \ln \lambda _{b}=\ln 2-(1+r)^{-1}\ln r \).
The solid line in Fig. \ref{fig: DvslambdaNumerical} shows a plot of \( D_{\Lambda } \)
versus \( \ln \lambda _{1} \) for \( r=3 \). Note that for large \( \lambda _{1} \),
\( D_{\Lambda }=D_{c} \) is independent of \( \lambda _{1} \).

It is also instructive to consider the case of uniform stretching \( (r=1) \)
for which \( \lambda _{1}=\lambda _{2} \). In that case, \( h_{a}=1/\tau  \),
and there is a rigorous known result for the dimension \cite{KM-PY}. For \( \lambda _{1}=\lambda _{2} \),
Eqs. (\ref{Da})--(\ref{Dc}) yield
\begin{equation}
\label{Dimension_Uniform2D}
D_{\Lambda }=\left\{ \begin{array}{ccl}
2-\frac{\ln \lambda _{1}}{\ln 2} & \textrm{for} & 1\leq \lambda _{1}\leq 2,\\
1 & \textrm{for} & \lambda _{1}\geq 2.
\end{array}\right. 
\end{equation}
(For \( r\rightarrow 1 \) region (b), where \( D_{\Lambda }=D_{b} \), shrinks
to zero width in \( \lambda _{1} \).) For \( r=1 \) the natural transient
measure is uniform; from (\ref{musm2D}) we have \( \mu (s^{(n)}_{m})=2^{-n} \)
independent of the interval (i.e, independent of \( m \)). In this case there
is no difference between the capacity dimension of the invariant set and the
information dimension of its measure. Equation (\ref{Dimension_Uniform2D})
agrees with the rigorous known result, thus lending support to the original
conjecture.

\subsection{Numerical Tests}

\begin{figure}
{\par\centering \resizebox*{0.9\columnwidth}{!}{\includegraphics{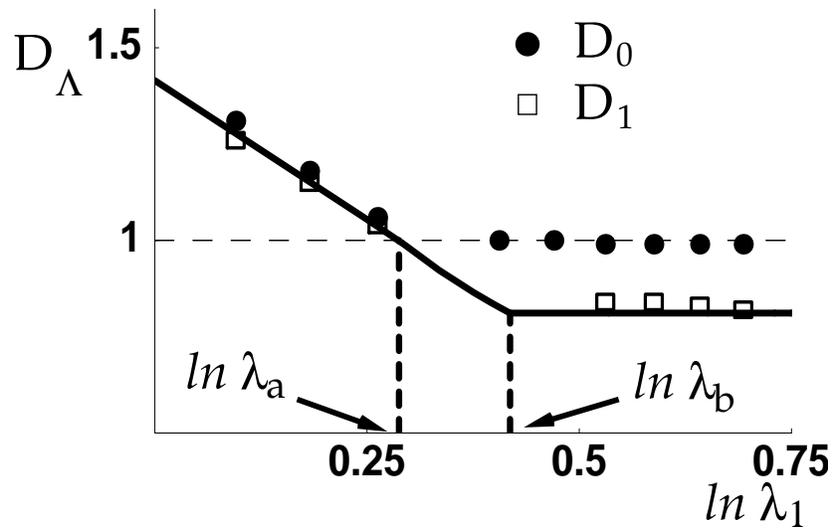}} \par}

\caption{\label{fig: DvslambdaNumerical}Comparison of prediction and numerical dimension
calculations for the 2D model map, (\ref{2DMap-1})-(\ref{2DMapa}), with \protect\( r=\lambda _{2}/\lambda _{1}=3\protect \)
fixed. The filled dots show calculations of the box-counting dimension, and
the boxes show information dimension. The solid line shows the predicted information
dimension as given by Eqs. (\ref{Da})-(\ref{Dc}).}
\end{figure}

The formulae (\ref{Da})--(\ref{Dc}) were verified by numerical measurements
of the information dimension, \( D_{1} \), of \( \Lambda  \) at various values
of \( \lambda _{1} \) with \( r=\lambda _{2}/\lambda _{1} \) fixed at \( r=3 \)
(Fig. \ref{fig: DvslambdaNumerical}). Shown for comparison is the box-counting
dimension, \( D_{0} \). The values of the box-counting dimension are numerically
indistinguishable from the values of the information dimension when \( D_{0},\textrm{ }D_{1}>1 \),
or \( \lambda _{1}<\lambda _{a} \), the region corresponding to formula (\ref{Da}).
For \( \lambda _{1}>\lambda _{a} \), \( \Lambda  \) is a smooth curve and
so has a box-counting dimension of \( D_{0}=1 \). No points for \( D_{1} \)
are shown near \( \lambda _{1}=\lambda _{b} \). It can be argued (Appendix
A) that numerical convergence is too slow here to yield accurate measurements
of the dimension.

To numerically determine the information dimension of \( \Lambda  \) (the data
shown as open squared in Fig. \ref{fig: DvslambdaNumerical}), we place a square
\( x-y \) grid with a spacing \( \varepsilon  \) between grid points over
a region containing \( \Lambda  \). Using the method dscribed in the next paragraph
we compute the natural measure in each grid box and repeat for various \( \varepsilon  \).
The information dimension is then given by 
\[
D_{1}=\lim _{\varepsilon \rightarrow 0}\frac{I(\varepsilon )}{\ln (1/\varepsilon )},\]
where \( I(\varepsilon )=\sum ^{N(\varepsilon )}_{i=1}\mu _{i}\ln (1/\mu _{i}) \)
is a sum over the \( N(\varepsilon ) \) grid boxes which intersect \( \Lambda  \)
and \( \mu _{i} \) is the natural measure in the \( i \)th box. The slope
of a plot of \( I(\varepsilon ) \) versus \( \ln \varepsilon  \) gives \( D_{1} \).
The box-counting dimension (the data shown as black dots in Fig. \ref{fig: DvslambdaNumerical})
is given by

\[
D_{0}=\lim _{\varepsilon \rightarrow 0}\frac{\ln N(\varepsilon )}{\ln 1/\varepsilon }\]
and calculated in an analogous way.

To determine which boxes intersect \( \Lambda  \) and what measure is contained
in each of them we take advantage of the fact that \( \Lambda  \) is a function
\cite{OttBook} {[}see Appendix A, Eq. (\ref{Lambda_function}){]}. That is,
for each value of \( x \) there is only one corresponding value of \( y \)
in \( \Lambda  \), which we denote \( y=y_{\Lambda }(x) \). We divide the
interval \( 0\leq x<1 \) into \( 2^{n} \) intervals of width \( \delta \equiv 2^{-n} \).
We wish to approximate \( y_{\Lambda }(x_{0}) \) for \( x_{0} \) in the center
of the \( x \)-interval. To do this we iterate \( x_{0} \) forward using (\ref{2DMap-1})
\( m \) times until the condition
\begin{equation}
\label{Close_Condition}
\frac{\delta }{2}\lambda _{1}^{m_{1}}\lambda ^{m_{2}}_{2}\geq 1
\end{equation}
is first met, where \( m_{1} \) (\( m_{2} \)) is the number of times the orbit
lands in \( 0\leq x<1/2 \) (\( 1/2\leq x<1 \)), and \( m_{1}+m_{2}=m \) (we
will see the reason for this condition below). All of the values, \( x_{i} \),
of the iterates are saved. Starting now from \( x_{m} \) and taking \( y_{m}=0 \)
we iterate backward \( m \) times. For \( \eta  \) small enough, \( \Lambda  \)
is contained in \( -1\leq y\leq 1 \), so the point (\( x_{m} \), \( y_{m}=0 \))
is within a distance \( 1 \) in the \( y \)-direction of \( \Lambda  \).
The \( m \) reverse iterations shrink the segment \( y_{m}\leq y\leq y_{\Lambda }(x_{m}) \)
by a factor \( \lambda ^{m_{1}}_{1}\lambda ^{m_{2}}_{2} \) so that \( \left| y_{0}-y_{\Lambda }(x_{0})\right| \leq \delta /2 \)
by condition (\ref{Close_Condition}). Thus, we have found a point \( y_{0} \)
that approximates \( y_{\Lambda }(x_{0}) \) to within \( \delta  \). Since
we will be using \( \epsilon  \) boxes with \( \epsilon \gg \delta  \), we
may regard \( y_{0} \) as being essentially equal to \( y_{\Lambda }(x_{0}) \).
The measure in the \( \delta  \) width interval containing \( x_{0} \) is
found by iterating \( x_{0} \) forward \( n \) times and using equation (\ref{musm2D}),
\( \mu =\frac{\lambda ^{n_{2}}_{1}\lambda ^{n_{1}}_{2}}{(\lambda _{1}+\lambda _{2})^{n}} \),
where \( n_{1} \) (\( n_{2} \)) is the number of times the orbit lands in
\( 0\leq x<1/2 \) (\( 1/2\leq x<1 \)), and \( n_{1}+n_{2}=n \). We associate
this measure with the point \( (x_{0},y_{0}) \) (with \( y_{0} \) found by
the above procedure). 

Note that for fractal \( y_{\Lambda }(x) \) the \( y \) interval occupied
by the curve \( y=y_{\Lambda }(x) \) in an \( x \) interval of width \( \delta \ll 1 \)
is of order \( \delta ^{D_{0}-1} \) which is large compared to \( \delta  \).
We now cover the region with new grids having successively larger spacing, \( \varepsilon _{i}=2^{i}\delta =2^{i-n} \),
and calculate \( I(\varepsilon _{i}) \) and \( N(\varepsilon _{i}) \) based
on the data taken from the first \( \delta  \)-grid. For \( i \) large enough,
such that the \( y \) extent of the curve \( y_{\Lambda }(x) \) in a typical
\( \delta  \) width interval is less than \( \epsilon _{i} \) (i.e., \( \epsilon _{i}\stackrel{{\sim }}{>}\delta ^{2-D_{0}} \)
or \( i\stackrel{{\sim }}{>}(D_{0}-1)n \)) we observe linear scaling of \( \log I(\epsilon _{i}) \)
and \( \log N(\epsilon _{i}) \) with \( \log \epsilon _{i} \), and we use
the slope of such plots to determine \( D_{1} \) and \( D_{0} \). The dimensions
\( D_{1} \) and \( D_{0} \) are then determined as described above.

\subsection{Atypical Case}

The conjecture of Ref. \cite{HOY} is that the dimension formulae of Sec. 2
apply for ``typical'' systems. To see the need for this restriction consider
Eqs. (\ref{2DMap-2}) for the case where \( \eta =0 \). It is easily shown
that the dimension formulae can be violated in this case. The claim, however,
is that \( \eta =0 \) is special, or ``atypical'', in that, as soon as we
give \( \eta  \) any nonzero value, the validity of the dimension formulae
is restored. In this connection it is important to note that as long as \( \eta \neq 0 \),
the dimension of the invariant set is independent of the value of \( \eta  \).
This follows since if \( \eta \neq 0 \) we can always rescale the value of
\( \eta  \) to one by the change of variables \( \tilde{y}=y/\eta  \). To
see the violation of the dimension formulae for \( \eta =0 \), we note that
in this case, by virtue of (\ref{2DMap-2}), the line \( y=0 \) is invariant.
Thus the measure is distributed on a one-dimensional subspace, the \( x \)-axis.
(This is very different from the picture in Fig. \ref{fig: 2D Fractal}, where
the invariant set, the boundary between black and white, appears to be fractal.)
Using the definition of the information dimension and dividing the \( x \)-axis
into intervals of width \( 2^{-n} \), the information dimension of the natural
measure is 
\begin{equation}
\label{3.17}
D_{\Lambda }=\lim _{n\rightarrow \infty }\frac{\sum ^{n-1}_{m=0}\mu (s^{(n)}_{m})\ln [1/\mu (s^{(n)}_{m})]}{\ln (2^{n})}.
\end{equation}
The quantity whose limit is taken in (\ref{3.17}) is in fact independent of
\( n \). Thus, taking \( n=1 \) we obtain for \( D_{\Lambda } \) the result
that, for \( \eta =0 \), 
\[
D_{\Lambda }=D_{c},\]
\emph{for all} \( \lambda _{1} \) and \( \lambda _{2}>1 \), where \( D_{c} \)
is given by (\ref{Dc}). Thus, for \( h_{a}<h_{b} \), \( D_{\Lambda } \) is
greater when \( \eta \neq 0 \) than when \( \eta =0 \), and, thus, the conjectured
stable manifold dimension formula of Sec. 2 is violated. For \( h_{a}>h_{b} \),
\( D_{\Lambda } \) is the same in both cases.

In Sec. 4 we consider chaotic scattering in a three-dimensional billiard example
for which the character of the atypical case is more interesting than in the
above example. In particular, we find for our billiard that in both the typical
and the atypical cases the stable manifold can have noninteger capacity dimension.

\subsection{General Considerations}

\begin{figure}
{\par\centering \resizebox*{0.9\columnwidth}{!}{\includegraphics{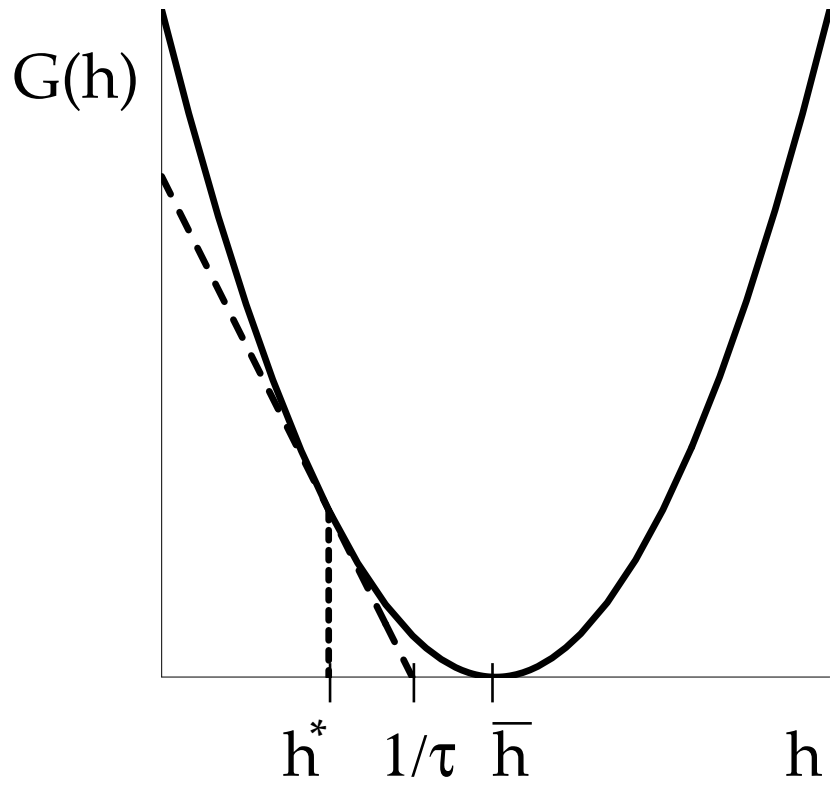}} \par}

\caption{\label{fig: Gvsh}\protect\( G(h)\protect \) versus \protect\( h\protect \).
The long-dashed line has slope \protect\( -1\protect \) and is tangent to the
graph of \protect\( G(h)\protect \) at the point \protect\( h=h_{*}\protect \).}
\end{figure}

The previous considerations readily generalize to the case of an arbitrary smooth
function \( \lambda (x)>1 \) and a general chaotic map, \( x_{n+1}=M(x_{n}) \),
which replaces (3.1). Consider the finite time vertical Lyapunov exponent,
\[
\tilde{h}(x,n)=\frac{1}{n}\sum ^{n}_{m=1}\ln \lambda (M^{m-1}(x))\]
computed for the initial condition \( x \). Choosing \( x \) randomly with
uniform probability distribution in the relevant basin for chaotic motion {[}e.g.,
\( x \) in \( [0,1] \) for Eq. (\ref{2DMap-1}){]}, \( \tilde{h}(x,n) \)
can be regarded as a random variable. Let \( \tilde{P}(h,n) \) denote its probability
distribution function. For large \( n \), we invoke large deviation theory
to write \( \tilde{P}(h,n) \) as \cite{OttBook} 
\[
\ln \tilde{P}(h,n)=-nG(h)+o(n),\]
or, more informally, 
\begin{equation}
\label{uniformProbDist}
\tilde{P}(h,n)\sim e^{-nG(h)},
\end{equation}
where the specific form of \( G(h) \) depends on \( M(x) \) and the specific
\( \lambda (x) \), and \( G(h) \) is convex, \( d^{2}G(h)/dh^{2}\geq 0 \).
For the normalization, \( \int \tilde{P}(h,n)dh=1 \), to hold for \( n\rightarrow \infty  \),
we have that 
\[
\min _{h}G(h)=0;\]
see Fig. \ref{fig: Gvsh}, where \( \bar{h} \) denotes the value of \( h \)
for which the above minimum is attained. As \( n\rightarrow \infty  \) we see
that \( \tilde{P} \) approaches a delta function, \( \delta (h-\bar{h}) \).
Thus, \( \bar{h} \) is the usual infinite time Lyapunov exponent for almost
all initial conditions with respect to Lebesgue measure in \( 0\leq x\leq 1 \).

As described above \( \tilde{P}(h,n) \) is the probability distribution of
\( h(x,n) \) for \( x \) chosen randomly with respect to a uniform distribution
in \( [0,1] \). We now ask what the probability distribution of \( h(x,n) \)
is for \( x \) chosen randomly with respect to the natural transient measure
for our expanding map, \( x_{n+1}=M(x_{n}) \) and (\ref{2DMap-2}). To answer
this question we proceed as before and consider an initial vertical line segment
\( |y|\leq K \) starting at \( x \) {[}with \( K>(\eta /2\pi )(\lambda _{\min }-1)^{-1} \),
\( \lambda _{\min }=\min _{x}\lambda (x)>1 \){]}. After \( n \) iterations,
this line segment lengthens by the factor \( \exp [n\tilde{h}(x,n)] \). Thus,
the fraction of the line still remaining in the strip \( |y|<K \) is \( \exp [-n\tilde{h}(x,n)] \).
Hence, the fraction of points sprinkled uniformly in the strip that still remains
after \( n \) iterates is
\begin{equation}
\label{3.21}
e^{-n/\tau }\sim \int e^{-nG(h)-nh}dh,
\end{equation}
and the probability distribution of finite time vertical Lyapunov exponents
for \( x \) chosen randomly with respect to the natural transient measure is
\begin{equation}
\label{3.22}
P(h,n)\sim \frac{e^{-nG(h)-nh}}{\int e^{-nG(h)-nh}dh}.
\end{equation}
Evaluating (\ref{3.21}) for large \( n \) we have \( \int e^{-n[G(h)+h]}dh\sim e^{-n[G(h_{*})+h_{*}]} \),
where \( \min [G(h)+h]=G(h_{*})+h_{*} \) and \( h_{*} \) is the solution of
\( dG(h_{*})/dh_{*}=-1 \). Thus ,
\begin{equation}
\label{tauFromG}
1/\tau =G(h_{*})+h_{*}.
\end{equation}
See the construction in Fig. \ref{fig: Gvsh}, in which the dotted line of slope
\( -1 \) is tangent to the graph of \( G(h) \) at the point \( h=h_{*} \).
The infinite time vertical Lyapunov exponent for the transient natural measure
is 
\begin{equation}
\label{3.24}
h_{a}=\int hP(h,n)dh.
\end{equation}
Using (\ref{3.22}) and again letting \( n \) be large (\ref{3.24}) yields
\( h_{a}=h_{*} \). Referring to Fig. \ref{fig: Gvsh} we see that 
\[
h_{a}\leq 1/\tau .\]
That is, Eq. (\ref{ha_lt_tauinv}) is valid for general \( M(x) \) and \( \lambda (x) \)
and not just for \( M(x) \) and \( \lambda (x) \) given by (\ref{2DMap-1})
and (\ref{2DMapa}).

\section{A Three-Dimensional Billiard Chaotic Scatterer}

\begin{figure}
{\par\centering \resizebox*{0.9\columnwidth}{!}{\includegraphics{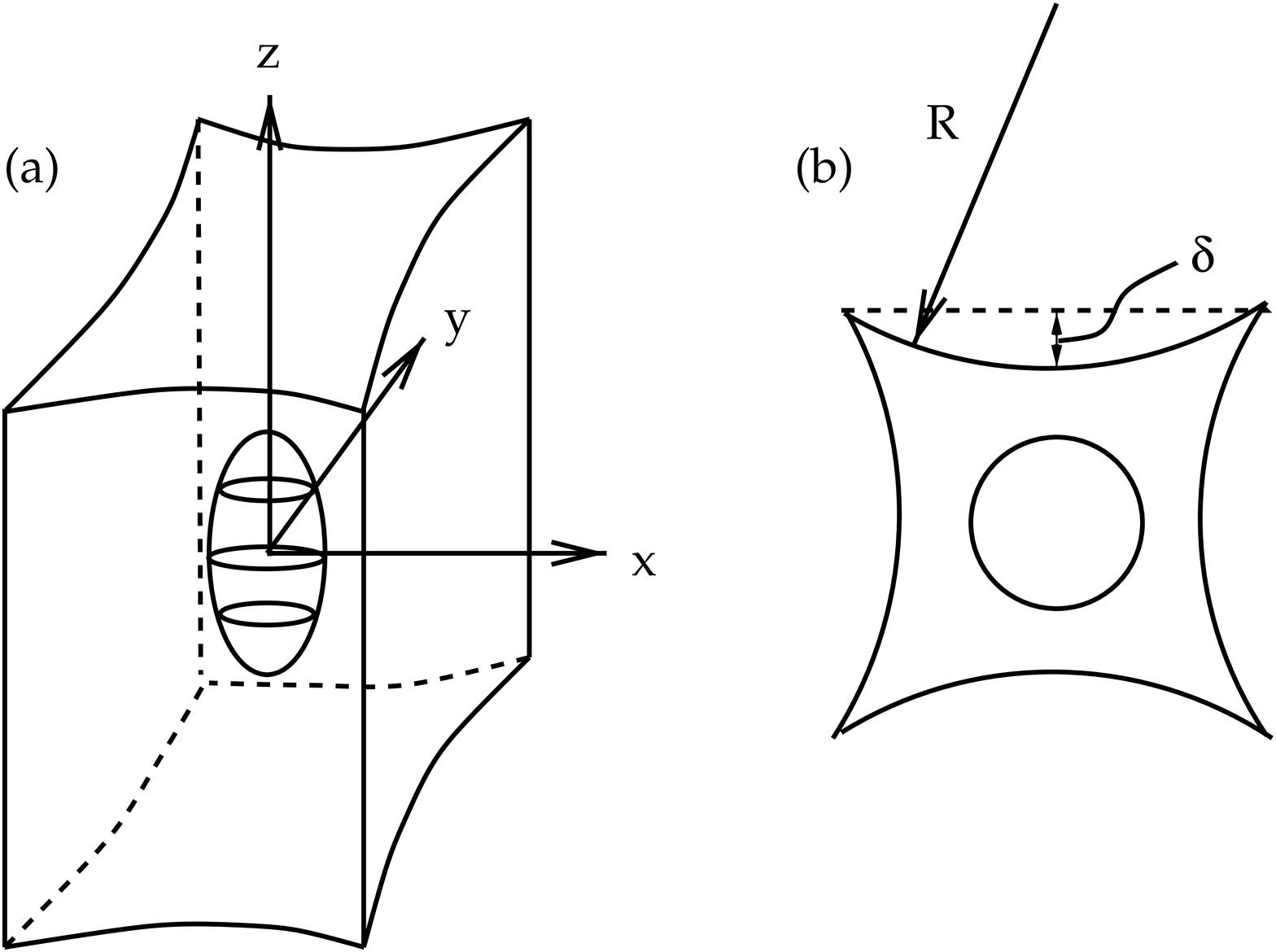}} \par}

\caption{\label{fig: EllInBox}The scattering system in the (atypical) case of vertical
orientation of the ellipsoid. (a) A hard ellipsoid is placed inside a hard tube
with cross-section as shown in (b). (The circle in (b) is the cross section
of the ellipsoid.) \protect\( R=25\protect \), \protect\( d=10\protect \)
and the radius of the ellipsoid at \protect\( z=0\protect \) is \protect\( r_{\parallel }=5\protect \). }
\end{figure}

We consider a three degree-of-freedom billiard. The billiard (Fig. \ref{fig: EllInBox})
is formed by a hard ellipsoid of revolution, placed in a hard, infinitely long
tube with cross-section as shown in Fig. \ref{fig: EllInBox}(b). The center
of the ellipsoid is placed at the center of the tube. We consider two cases:
(a) the major axis of the ellipsoid coincides with the \( z \)-axis {[}see
Fig \ref{fig: EllInBox}(a){]}, (b) the major axis of the ellipsoid lies in
the \( y \)-\( z \) plane and makes an angle \( \xi  \) with the \( z \)-axis.
The ratio of the minor radius of the ellipsoid (\( r_{\parallel } \)) to the
width of a side of the tube {[}dashed line in Fig. \ref{fig: EllInBox}(b){]}
is \( 1/4 \). This leaves the major radius, \( r_{\perp } \), and the tilt
angle, \( \xi  \), as parameters. A point particle injected into the system
experiences specular reflection from the ellipsoid and the walls (i.e., the
angle of reflection is equal to the angle of incidence, where both are taken
with respect to the normal to the surface off of which the particle bounces).
When the orbit has passed the top (bottom) of the ellipsoid, with positive (negative)
\( z \)-velocity, we say that it has exited upward (downward). We fix the conserved
energy so that \( \left| \vec{v}\right| =1 \).

\subsection{Pictures of the Stable Manifold}

We claim that the case of the untilted ellipsoid {[}\( \xi =0 \) and case (a)
above{]} is atypical in the sense that it violates the formulae of Sec. 2. However,
as soon as \( \xi \neq 0 \), we claim that the formulae of Sec. 2 apply. We
begin by discussing the atypical (untilted) case.

By the symmetry of the geometry of the billiard with the ellipsoid axis along
\( z \) {[}Fig. \ref{fig: EllInBox}(a){]}, the chaotic saddle, \( \Lambda  \),
of this system is the collection of initial conditions satisfying \( z=v_{z}=0 \).
Started with these initial conditions, an orbit will have \( z=v_{z}=0 \) for
all forward and reverse time. The surface normals of the ellipsoid and walls
at \( z=0 \) lie in the \( z=0 \) plane, and thus the particle cannot acquire
a non-zero \( v_{z} \). The \( z=0 \) slice through the three-dimensional
billiard is a two-dimensional billiard with concave walls {[}Fig. \ref{fig: EllInBox}(b){]}.
It is known \cite{OttBook} that a typical orbit in this billiard will fill
the phase space ergodically. Near \( z=0 \), we can picture a typical point
on the stable manifold (denoted \( SM \)) as having, for example, \( z \)
slightly less than zero and \( v_{z} \) slightly greater than zero. The particle
will hit the ellipsoid below its equator and, thus, \( v_{z} \) will be decreased
with each bounce, yet remain positive. With successive bounces, the orbit on
\( SM \) slowly approaches \( z=v_{z}=0 \), the chaotic saddle.

\begin{figure}
{\par\centering \resizebox*{0.9\columnwidth}{!}{\includegraphics{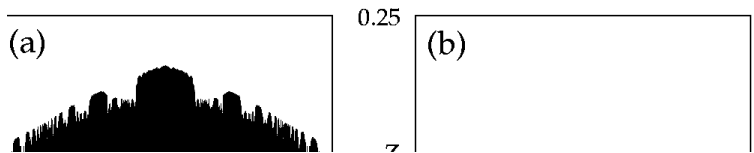}} \par}

\caption{\label{fig: Basin Boundaries (EB)}Basins for \protect\( z\rightarrow \infty \protect \)
(white) and \protect\( z\rightarrow -\infty \protect \) (black). \protect\( r_{\perp }=7\protect \),
\protect\( y=5.1\protect \), 4\protect\( v_{x}=0\protect \), \protect\( v_{z}=.1\protect \),
and \protect\( v_{y}\protect \) is given by the condition \protect\( \left| \vec{v}\right| =1\protect \)
for (a) the untilted case and (b) a tilt of \protect\( \frac{2\pi }{100}\protect \).}
\end{figure}

To visualize \( SM \), we note that it forms the boundary between initial conditions
which escape upward and those which escape downward. We say that points which,
when iterated, eventually escape upward (downward) are in the \emph{basin} of
upward (downward) escape. Points which are on the boundary between the two basins
never escape at all, i.e. they are in \( SM \). We initiate a (two-dimensional)
grid of orbits (500x500) on the plane,\( -3<x<3 \), \( y=5.1 \), \( -2.5<z<0 \),
\( v_{x}=0 \), \( v_{z}=.1 \), and \( v_{y} \) is given by the condition
\( \left| \vec{v}\right| =1 \). We iterate each of these initial conditions
forward until it escapes. Then we plot a white (black) point for each orbit
in the upward (downward) basin. The result is shown in Fig. \ref{fig: Basin Boundaries (EB)}(a).
The boundary between the white and black regions is then the intersection of
\( SM \) lying in the phase space of the five-dimensional billiard (\( x,y,z,v_{x},v_{y},v_{z} \)
constrained by \( \left| \vec{v}\right| =1 \)) with the specified two-dimensional
\( x,z \)-plane. \( SM \) appears to take the form of a nowhere-differentiable
curve. This is true in various 2D slices, none of which are chosen specially,
which suggests that \( SM \) has this form in a typical slice. A similar procedure
can be followed for the case of the tilted ellipsoid and the resulting picture
is shown in Fig. \ref{fig: Basin Boundaries (EB)}(b) for the case of a tilt
angle of \( \xi =2\pi /100 \) radians.

\subsection{Lyapunov Exponents, Decay Times, and Approximate Formulae for the Stable Manifold}

Again, we begin with the untilted case. To construct a map from this system
we record the cylindrical coordinates (\( z \), \( \phi  \)) and their corresponding
\( z \) and \( \phi  \) velocity components, which we denote (\( v \), \( \omega  \)),
each time the particle hits the ellipsoid. The coordinate \( r \) is constrained,
for a given \( z \), by the shape of the ellipsoid surface, and \( v_{r} \)
is given by the energy conservation condition \( \left| \vec{v}\right| =1 \).
The four components \( ( \)\( z \), \( v \), \( \phi  \), \( \omega  \)\( )_{n} \)
give the state of the system at discrete time \( n \), where \( n \) labels
the number of bounces from the ellipsoid. Let \( \vec{z}\equiv \left[ \begin{array}{c}
z\\
v
\end{array}\right]  \) and \( \vec{\phi }\equiv \left[ \begin{array}{c}
\phi \\
\omega 
\end{array}\right]  \). For future reference, we express the map using the following notation,
\begin{equation}
\label{Map}
\begin{array}{rcl}
\vec{z}_{n+1} & = & M_{z}(\vec{z}_{n},\vec{\phi }_{n}),\\
\vec{\phi }_{n+1} & = & M_{\phi }(\vec{z}_{n},\vec{\phi }_{n}).
\end{array}
\end{equation}
In what follows, when we refer to an orbit, saddle, invariant set, stable manifold,
etc., we are referring to these quantities for the discrete time map (rather
than the original continuous time system).

In the case of the untilted ellipsoid, linearizing about an orbit on \( \Lambda  \),
(i.e., \( \vec{z}_{n}=0 \), \( \vec{\phi }_{n} \)), we obtain, for the evolution
of differential orbit perturbations \( \delta \vec{z} \) and \( \delta \vec{\phi } \),
\[
\begin{array}{rcl}
\delta \vec{z}_{n+1} & = & DM_{z}(0,\vec{\phi }_{n})\delta \vec{z}_{n},\\
\delta \vec{\phi }_{n+1} & = & DM_{\phi }(0,\vec{\phi }_{n})\delta \vec{\phi }_{n},
\end{array}\]
where \( \vec{\phi }_{n+1}=M_{\phi }(0,\vec{\phi }_{n}) \) is the map for the
two-dimensional billiard of Fig. \ref{fig: EllInBox}(b), \( DM_{z}(0,\vec{\phi }) \)
is the tangent map for differential orbit perturbations in \( \vec{z} \) evaluated
at \( \vec{z}=0 \), and \( DM_{\phi }(0,\vec{\phi }) \) is the tangent map
for differential perturbations lying in \( \Lambda  \). Let \( \pm h_{z} \)
and \( \pm h_{\phi } \) denote the Lyapunov exponents with respect to the \emph{natural
transient measure} (Sec. 2) for perturbations in \( \vec{z} \) and in \( \vec{\phi } \),
respectively (these exponents occur in positive-negative pairs due to the Hamiltonian
nature of the problem). 

In principle, one could numerically evaluate \( h_{z} \) and \( h_{\phi } \)
by sprinkling a large number, \( N \), of initial conditions in the vicinity
of \( \Lambda  \), iterating \( n\gg 1 \) times, evaluating \( h_{z} \) and
\( h_{\phi } \) over those orbits still near \( \Lambda  \), and averaging
\( h_{z} \) and \( h_{\phi } \) over those orbits. We could also find \( \tau  \)
by this procedure; it is the exponential rate of decay of the orbits from the
vicinity of \( \Lambda  \). For cases where the escape time \( \tau  \) is
not long, this procedure, however, becomes problematic. For finite \( N \)
the number of retained orbits can be small or zero if \( n \) is too large.
Thus, we adopt an alternate procedure which we found to be less numerically
demanding.

In particular we make user of the ideas presented in Sec. 3.5. We define the
\emph{uniform measure} as the measure generated by uniformly sprinkling many
initial conditions in \( \Lambda  \) (the hyperplane \( z=v_{z}=0 \)). An
average over these orbits of the tangent space stretching exponents would yield
uniform measure Lyapunov exponents. We denote the distribution of finite-time
Lyapunov exponents with respect to this measure by \( P(\tilde{h}_{\phi },\tilde{h}_{z},n)\sim e^{-nG(\tilde{h}_{\phi },\tilde{h}_{z})} \)
{[}as in Eq. (\ref{uniformProbDist}){]}, where the tilde indicates finite time
exponents for initial conditions distributed according to the uniform measure.

To compute the decay time and the Lyapunov exponents with respect to the natural
transient measure we note that orbits near a point \( \vec{\phi } \) in \( \Lambda  \)
iterate away from \( \Lambda  \) as \( \exp [n\tilde{h}_{z}(\vec{\phi },n)] \).
Thus, the fraction of a large number of initial conditions sprinkled near \( \Lambda  \)
which remain near \( \Lambda  \) after \( n \) iterates is
\[
\int P(\tilde{h}_{\phi },\tilde{h}_{z},n)e^{-n\tilde{h}_{z}}d\tilde{h}_{z},\]
and \( h_{\phi } \) (the infinite time Lypunov exponent with respect to the
natural transient measure) is
\[
h_{\phi }=\lim _{n\rightarrow \infty }\frac{\int \tilde{h}_{\phi }P(\tilde{h}_{\phi },\tilde{h}_{z},n)e^{-n\tilde{h}_{z}}dh}{\int P(\tilde{h}_{\phi },\tilde{h}_{z},n)e^{-n\tilde{h}_{z}}dh}.\]
This expression can be approximated numerically by choosing \( N \) initial
conditions, \( \vec{\phi }_{i} \), uniformly in \( \Lambda  \) and calculating
\[
\left\langle n\tilde{h}_{\phi }\right\rangle _{n}\equiv \frac{n\sum ^{N}_{i=1}\tilde{h}_{\phi }(\vec{\phi }_{i},n)e^{-n\tilde{h}_{z}(\vec{\phi }_{i},n)}}{N\sum ^{N}_{i=1}e^{-n\tilde{h}_{z}(\vec{\phi }_{i},n)}}.\]
We calculate the finite-time Lyapunov exponents \( \tilde{h}_{\phi }(\vec{\phi }_{i},n) \)
and \( \tilde{h}_{z}(\vec{\phi }_{i},n) \) using the QR decomposition method
\cite{A}. Since we chose the \( \vec{\phi }_{i} \) uniformly in \( \Lambda  \)
these exponents are distributed according to \( P(n,\tilde{h}_{\phi },\tilde{h}_{z}) \).
\( N \) is taken to be large enough that there are at least \( 100 \) terms
contributing to \( 90\% \) of each sum. The range of \( n \) is from \( 10 \)
to about \( 40 \). We find that \( \left\langle n\tilde{h}_{\phi }\right\rangle _{n} \)
versus \( n \) is well-fitted by a straight line and we take its slope as our
estimate of \( h_{\phi } \). 

To find \( \tau  \), first note that we numerically find that \( h_{\phi } \)
does not vary much from \( \bar{h}_{\phi } \) (\( .91\leq h_{\phi }/\bar{h}_{\phi }\leq 1 \))
as we change the system parameter (the height of the ellipsoid), where the overbar
denotes an infinite-time Lypunov exponent with respect to the uniform measure
on \( \Lambda  \). Thus, we make the approximation \( P(\tilde{h}_{\phi },\tilde{h}_{z},n)\approx P(\tilde{h}_{z},n) \),
\( G(\tilde{h}_{\phi },\tilde{h}_{z})\approx G(\tilde{h}_{z}) \) and plot \( G(\tilde{h}_{z}) \)
versus \( \tilde{h}_{z} \). The value of \( 1/\tau  \) is given by Eq. (\ref{tauFromG}),
and \( h_{z} \) is given by \( dG(n,h_{z})/d\tilde{h}_{z}=-1 \). (Note that
\( h_{z}\leq 1/\tau \leq \bar{h}_{z} \).) A third order polynomial is fit to
the data for \( G(\tilde{h}_{z}) \) and used to find \( \tau  \) and \( h_{z} \).
This calculation is performed at a value of \( n \) which allows a significant
number of points in \( G(\tilde{h}_{z}) \) versus \( \tilde{h}_{z} \) to be
collected near in the range \( h_{z}<\tilde{h}_{z}<\bar{h}_{z} \). 

For the case of the tilted ellipsoid, we will consider a very small tilt angle,
\( \xi =2\pi /100 \). With this small tilt angle, the Lyapunov exponents and
decay times for the tilted and the untilted cases are approximately the same.
Thus, for the tilted case, we will use the same Lyapunov exponent values, \( \pm h_{z} \)
and \( \pm h_{\phi } \) and decay time \( \tau  \), that we numerically calculated
for the untilted case. The dimension formula of Sec. 2 for \( SM \) is
\begin{equation}
\label{PresentTilted}
\begin{array}{cc}
D_{S}=4-(h_{\phi }\tau )^{-1} & \, \textrm{for}\, h_{\phi }\tau \geq 1,\\
D_{S}<3 & \, \textrm{for}\, h_{\phi }\tau <1.
\end{array}
\end{equation}
(As for cases (b) and (c) of Sec. 3.2, we could write explicit expressions for
\( D_{S} \) for \( h_{\phi }\tau <1 \), but, in what follows we only need
that \( D_{S}<3 \) in this case.)

If the tilt angle is made to be zero (\( \xi =0 \)), we find that \( D_{S} \)
is not given by (\ref{PresentTilted}), but by the following formula (which
we derive in Sec. 7.2),
\begin{equation}
\label{PresentUntilted}
\begin{array}{cc}
D_{S}=4-\frac{h_{z}+1/\tau }{h_{\phi }} & \, \textrm{for}\, h_{\phi }\geq h_{z}+1/\tau ,\\
D_{S}<3 & \, \textrm{for}\, h_{\phi }<h_{z}+1/\tau .
\end{array}
\end{equation}
Since \( D_{S} \) is given by (\ref{PresentUntilted}) only if the tilt angle
\( \xi  \) is precisely zero, we say that the untilted ellipsoid scattering
system is atypical. As conjectured in \cite{HOY}, \( D_{S} \) from (\ref{PresentTilted})
is greater than or equal to \( D_{S} \) from (\ref{PresentUntilted}). Note
that \( D_{S} \) from the first line of (\ref{PresentTilted}) is larger that
\( D_{S} \) from the first line of (\ref{PresentUntilted}) by the factor \( h_{z}/h_{\phi } \).
Although the transition of \( D_{S} \) from \( \xi =0 \) to \( \xi \neq 0 \)
is strictly discontinuous, there is also a continuous aspect: In numerically
calculating the dimension of a measure one typically plots \( \ln I(\epsilon ) \)
versus \( \ln (1/\epsilon ) \), where \( I(\epsilon )=\sum \mu _{i}\ln [1/\mu _{i}] \)
and \( \mu _{i} \) is the measure of the \( i^{\textrm{th}} \) cube in an
\( \epsilon  \) grid. One then estimates the dimension as the slope of a line
fitted to small \( \epsilon  \) values in such a plot. In the case of very
small tilt, such a plot is expected to yield a slope given by (\ref{PresentUntilted})
for \( \epsilon \stackrel{{\sim }}{>}\epsilon _{*} \) and subsequently, for
\( \epsilon \stackrel{{\sim }}{<}\epsilon _{*} \), to yield a slope given by
(\ref{PresentTilted}). Here \( \epsilon _{*} \) is a small tilt-dependent
cross-over value, where \( \epsilon _{*}\rightarrow 0 \) as \( \xi \rightarrow 0 \).
In such a case, the dimension, which is defined by the \( \epsilon \rightarrow 0 \)
limit, is given by (\ref{PresentTilted}).

\section{Numerical Computations for the Three-Dimensional Billiard Scatterer}

\begin{figure}
{\par\centering \resizebox*{0.9\columnwidth}{!}{\includegraphics{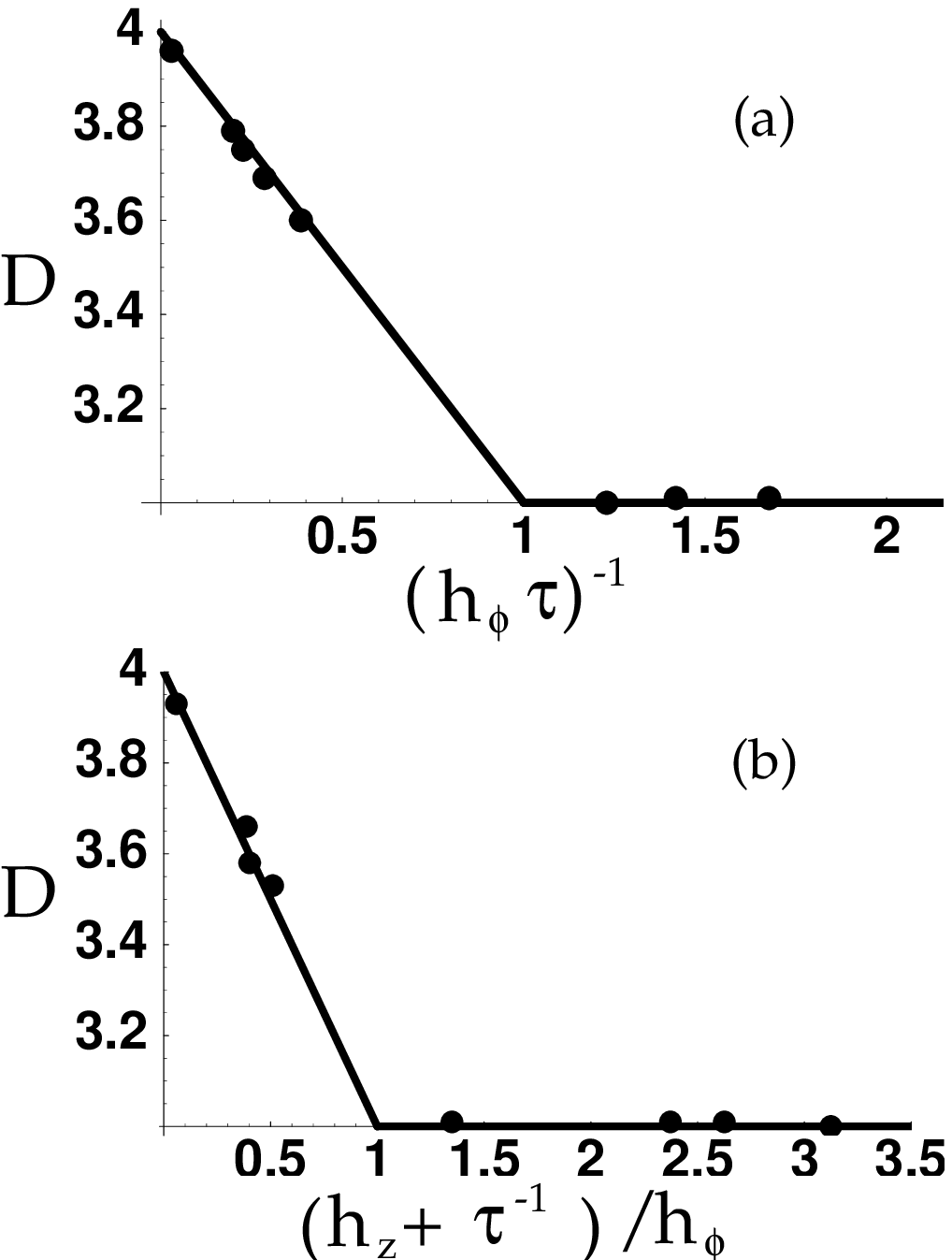}} \par}

\caption{\label{fig: EBGraph}Comparison of dimension formulae, Eq. (\ref{PresentTilted})
and (\ref{PresentUntilted}), with numerical estimates. The lines are the graphs
of the equations and the dots show the numerical estimates of the box-counting
dimension. (a) The tilted ellipsoid (typical) case. (b) The untilted ellipsoid
(atypical) case.}
\end{figure}

To verify that the untilted system is atypical we numerically calculated the
box-counting dimension of \( SM \) for various values of the parameter \( r_{\perp } \),
then introduced a small tilt perturbation in the form of a \( -2\pi /100 \)
radian tilt of the ellipsoid about the \( x \)-axis and repeated the dimension
calculations. The results are shown in Fig. \ref{fig: EBGraph}. They confirm
Eqs. (\ref{PresentTilted}) and (\ref{PresentUntilted}). (We do not plot data
points near the kink in the theoretical curve because of the numerical difficulty
in obtaining reliable results in that parameter range. This is a result of slow
converge near the kink. It is also found in the example of Sec. 3.3 and discussed
in Appendix A.)

We compare measured values of the box-counting dimension to the predicted values
of the information dimension in Fig. \ref{fig: EBGraph}. The box-counting dimension
gives an upper bound on the information dimension, but often the values of the
two dimensions are very close. In particular, we can compare the result for
the tilted ellipsoid system to the 2D map of Sec. 3. In the regime \( h_{\phi }\geq 1/\tau  \),
our system is similar to case (a) studied in Section 3.2 (note that, as discussed
in Sec. 4, \( 1/\tau  \) is never smaller than \( h_{z} \)). Changing \( r_{\perp } \)
while leaving \( r_{\parallel } \) fixed changes \( \tau  \) while \( h_{\phi } \)
changes only slightly. This is similar to varying \( \lambda _{1} \) of the
2D map while keeping \( \lambda _{2} \) fixed, i.e. varying \( r \). Figure
\ref{fig: DvslambdaNumerical}, shows (for a particular value of \( r \)) that
the measured values of \( D_{0} \) and \( D_{1} \) are numerically indistinguishable
in the region corresponding to case (a), \( \lambda <\lambda _{a} \). 

For \( h_{\phi }\tau <1 \) (\( h_{\phi }<h_{z}+1/\tau  \)) the information
dimension of \( SM \) is predicted to be less than three for the tilted (untilted)
ellipsoid system. This is analogous to cases (b) and (c) of Sec. 3.2; see Fig.
\ref{fig: DvslambdaNumerical}. Since \( SM \) divides the four-dimensional
phase space, its box-counting dimension cannot be less than three, and, analogous
to the result of Sec. 3 and Fig. \ref{fig: DvslambdaNumerical}, we expect that
the box-counting dimension is three for \( h_{\phi }\tau <1 \) (\( h_{\phi }<h_{z}+1/\tau  \)).
This expectation is borne out by the numerical results, Fig. \ref{fig: EBGraph}.

The box-counting dimension of \( SM \) was computed using the uncertainty dimension
method \cite{McDGY,OttBook}. This method gives the box-counting dimension of
the basin boundary which, as discussed above, coincides with \( SM \). 

The uncertainty dimension method was carried out as follows:

\begin{enumerate}
\item Choose a point, \( \vec{x} \), at random in a region of a 2D plane intersecting
the basin boundary (e.g., Fig. 6) and determine by iteration in which basin
it lies.
\item Determine in which basins the perturbed initial points \( \vec{x}\pm \vec{\delta } \)
lie (\( \vec{\delta } \) is some small vector).
\item If the three points examined in (1) and (2) do not all lie in the same basin,
then \( \vec{x} \) is called ``uncertain''. 
\item Repeat 1 to 3 for many points \( \vec{x} \) randomly chosen in the 2D plane,
and obtain the fraction of these that are uncertain.
\item The fraction of points which is uncertain for a given \( \vec{\delta } \),
denoted \( f \), scales like \cite{PELIKAN} \( f\sim |\vec{\delta }|^{2-d_{0}}, \)
where \( d_{0} \) is the box-counting dimension of the intersection of \( SM \)
with the 2D plane. The box-counting dimension of \( SM \) in the full 4D state
space of the map is \( D_{0}=2+d_{0} \). (The dimension of a generic intersection
of a 2D plane with a set having dimension \( D_{0} \) in a 4D space is \( d_{0}=2+D_{0}-4 \),
which gives \( D_{0}=2+d_{0} \).) Thus, plot \( \ln f \) versus \( \ln \left| \vec{\delta }\right|  \),
fit a straight line to the plot, and estimate \( D_{0} \) as \( 4 \) minus
the slope of this line. See Fig. \ref{fig: fvseGraph} for an example of such
a plot.
\end{enumerate}

\section{Structure of the Stable Manifold}

\subsection{Untilted Ellipsoid}

Due to the symmetry to the untilted ellipsoid billiard, the chaotic saddle of
the untilted ellipsoid system has a special geometry (i.e., it lies in \( z=v=0 \)),
which, as we show, accounts for the dimension being lower than the predicted
value for a typical system. Similarly, the symmetry induces a special geometry
on the stable manifold (\( SM \)). Figure \ref{fig: BB3D} shows a 3D slice
of \( SM \) in the atypical system. The slice is at a fixed value of \( \omega  \).
The axes are \( z \), \( v \), and \( \phi  \), but one could have chosen
an arbitrary line through (\( \phi  \),\( \omega  \)) as the third axes and
seen a plot which was qualitatively the same. The stable manifold is organized
into rays emanating from the \( \phi  \)-axis with oscillations along the \( \phi  \)-direction.
The magnitude of the oscillations decreases to zero as the \( \phi  \)-axis
is approached.

To understand the structure of \( SM \) in more detail, assume that \( \left| \vec{z}\right|  \)
is small. Then, since \( \left| \vec{z}\right| =0 \) is invariant, we can approximate
the dynamics by expanding to first order in \( \vec{z} \),
\begin{equation}
\label{ZEqnAgain}
\vec{z}_{n+1}\cong DM_{z}(0,\vec{\phi }_{n})\vec{z}_{n},
\end{equation}
\begin{equation}
\label{PhiEqnAgain}
\vec{\phi }_{n+1}\cong M_{\phi }(0,\vec{\phi }_{n}).
\end{equation}
Say \( (\vec{z}_{SM},\vec{\phi }_{SM}) \) is a point on \( SM \). As this
point is iterated we have that \( \left| \vec{z}\right| \rightarrow 0 \) with
increasing \( n \). However, since (\ref{ZEqnAgain}) is linear in \( \vec{z}_{n} \),
for any constant \( \alpha  \), and the initial condition, \( (\alpha \vec{z}_{SM},\vec{\phi }_{SM}) \),
the subsequent orbit must also have \( \left| \vec{z}\right| \rightarrow 0 \).
Consequently, if \( (\vec{z}_{SM},\vec{\phi }_{SM}) \) lies in \( SM \), so
does \( (\alpha \vec{z}_{SM},\vec{\phi }_{SM}) \). Thus, for the system (\ref{ZEqnAgain}),
(\ref{PhiEqnAgain}), the stable manifold at any point \( \vec{\phi } \) lies
on a straight line through the origin of the two-dimensional \( \vec{z} \)-space.
Put another way, in the approximation (\ref{ZEqnAgain}), (\ref{PhiEqnAgain}),
the stable manifold can be specified by an equation giving the \emph{angle}
of \( \vec{z} \) as a function of \( \vec{\phi } \). Thus, decomposing into
polar coordinates \( (\rho ,\chi ) \), the stable manifold for \( \left| \vec{z}\right| \rightarrow 0 \)
approaches the form 
\begin{equation}
\label{ChiOfPhi}
\chi =\chi (\vec{\phi }).
\end{equation}
(This explains the structure seen in Fig. \ref{fig: BB3D}.) For \( \left| \vec{z}\right|  \)
finite the linearity of (\ref{ZEqnAgain}) is not exact, and we expect that
the behavior of the stable manifold at constant \( \vec{\phi } \) is not a
straight line through the origin of the \( \vec{z} \) plane. Rather, as \( \left| \vec{z}\right|  \)
becomes larger we expect (and numerically observe) the straight line for small
\( \left| \vec{z}\right|  \) to appear as a smooth curve through \( \vec{z}=0 \). 

Since for fixed \( \vec{\phi } \) \( SM \) varies smoothly with increasing
\( \rho =\left| \vec{z}\right|  \), the dimension of \( SM \) is not affected
by the approximation (\ref{ZEqnAgain}) and (\ref{PhiEqnAgain}). That is, to
find the dimension of \( SM \), we can attempt to find it in the region of
small \( \left| \vec{z}\right|  \) where (\ref{ZEqnAgain}) and (\ref{PhiEqnAgain})
are valid, and that determination will apply to the whole of \( SM \).

The task of analytically determining \( D_{S} \) is too hard for us to accomplish
in a rigorous way for the system, (\ref{ZEqnAgain}), (\ref{PhiEqnAgain}),
applying to our billiard {[}a heuristic analysis yielding (\ref{PresentUntilted})
is given in Sec. 7.2{]}. Thus, to make progress, we adopt a model system with
the same structure as (\ref{ZEqnAgain}), (\ref{PhiEqnAgain}). In particular,
we wish to replace the two-dimensional billiard map (\ref{PhiEqnAgain}) by
a simpler map, \( M_{\phi }\rightarrow \bar{M}_{\phi } \), that, like the original
two-dimensional billiard map, is chaotic and describes a Hamiltonian system.
For this purpose we choose the cat map,
\begin{equation}
\label{analPhiMap}
\vec{\phi }_{n+1}=\bar{M}_{\phi }(\vec{\phi }_{n})\equiv C\vec{\phi }_{n}\, \textrm{modulo}\, 1,
\end{equation}
where \( C \) is the cat map matrix,
\[
C=\left[ \begin{array}{cc}
2 & 1\\
1 & 1
\end{array}\right] .\]
Similarly, we replace \( DM_{z}(0,\vec{\phi }) \) in (\ref{ZEqnAgain}) by
a simple symplectic map depending on \( \vec{\phi } \),
\begin{equation}
\label{analZMap}
\vec{z}_{n+1}=\bar{M}_{z}(\vec{z}_{n})=\left[ \begin{array}{cc}
\lambda  & f(\vec{\phi }_{n})\\
0 & \lambda ^{-1}
\end{array}\right] \vec{z}_{n},
\end{equation}
where \( \lambda >1 \), and \( f(\vec{\phi }) \) is a smooth periodic function
with period one in \( \phi  \) and \( \pi  \). {[}Note that vertical (i.e.,
parallel to \( z \)) line segments are uniformly expanded by the factor \( \lambda  \),
and thus, by the same argument as in Sec. 3.1, we have \( 1/\tau =\ln \lambda  \)
and \( h_{z}=1/\tau  \).{]}

Since only the angle of \( \vec{z} \) is needed to specifiy the stable manifold,
we introduce the variable \( \nu =z/v=\tan \chi  \). We can then derive a map
for \( \nu  \): From (\ref{analZMap}) we have \( \nu _{n+1}v_{n+1}=\lambda \nu _{n}v_{n}+v_{n}f(\vec{\phi }_{n}) \)
and \( v_{n+1}=\lambda ^{-1}v_{n} \). Dividing the first equation by the second,
\( v_{n+1} \) and \( v_{n} \) cancel and we obtain 
\begin{equation}
\label{analNuMap}
\nu _{n+1}=\lambda ^{2}\nu _{n}+\lambda f(\vec{\phi }).
\end{equation}
We now consider the dynamical system consisting of (\ref{analPhiMap}) and (\ref{analNuMap}).
Note that the system (\ref{analPhiMap}), (\ref{analNuMap}) is a three-dimensional
map, unlike the system (\ref{analPhiMap}), (\ref{analZMap}), which is a four
-dimensional map.

For (\ref{analPhiMap}), (\ref{analNuMap}), the stable manifold is given by
\begin{equation}
\label{ModelSM}
\nu =\nu _{S}(\vec{\phi })=-\lambda ^{-1}\sum ^{\infty }_{i=0}\lambda ^{-2i}f(C^{i}\vec{\phi }_{0}),
\end{equation}
where \( C \) is the cat map matrix. 

To verify that this is \( SM \) we note that points above \( SM \), \( \nu >\nu _{S}(\vec{\phi }) \)
{[}below \( SM \), \( \nu <\nu _{S}(\vec{\phi }) \){]} are repelled toward
\( \nu \rightarrow \infty  \) (\( \nu \rightarrow -\infty  \)). Thus, on backwards
iteration, points go toward \( SM \). We take advantage of this behavior to
determine \( SM \). Imagine that we iterate \( \vec{\phi } \) forward \( n \)
iterates to \( \vec{\phi }_{n}=C^{n}\vec{\phi }\, \textrm{modulo}\, 1 \), then
choose a value of \( \nu _{n} \), and iterate it backwards using (\ref{analNuMap}).
We find that the initial value of \( \nu  \) at time zero giving the chosen
value \( \nu _{n} \) at time \( n \) is
\[
\nu _{0}=(\nu _{n}/\lambda ^{2n})-\lambda ^{-1}\sum ^{n-1}_{i=0}\lambda ^{-2i}f(C^{i}\vec{\phi }).\]
Keeping \( \nu _{n} \) fixed and letting \( n\rightarrow +\infty  \), the
value of \( \nu _{0} \) approaches \( \nu _{S}(\vec{\phi }) \), given by (\ref{ModelSM}).

Results proven in \cite{KM-PY} show that the box-counting dimension (the capacity)
of the graph of the function \( \nu =\nu _{S}(\vec{\phi }) \) given by (\ref{ModelSM})
is
\begin{equation}
\label{DHatFromKM-PY}
\hat{D}_{S}=\left\{ \begin{array}{cc}
3-2\frac{\ln \lambda }{\ln B}, & \textrm{for }\lambda \leq B,\\
2, & \textrm{for }\lambda >B,
\end{array}\right. 
\end{equation}
where \( B=\frac{3+\sqrt{5}}{2}>1 \) is the larger eigenvalue of the matirx
\( C \). The formula for \( \lambda <B \) holds for almost all (with respect
to Lebesgue measure) values of \( \lambda  \). Since \( \lambda >1 \), the
sum in (\ref{ModelSM}) converges absolutely, implying that \( \nu _{S}(\vec{\phi }) \)
is a continuous function of \( \vec{\phi } \). Thus, when the first result
in Eq. (\ref{DHatFromKM-PY}) applies (i.e., the surface is fractal with \( \hat{D}_{S}>2 \)),
the stable manifold is a continuous nondifferentiable surface.

For example, evaluating (\ref{ModelSM}) on the surface \( \vec{\phi }=s\hat{u}_{+} \)
where \( \hat{u}_{+} \) is the unit vector in the eigendirection of \( C \)
corresponding to the expanding eigenvalue \( B \), we have that \( \nu  \)
versus \( s \) is of the form
\[
\nu =-\sum ^{\infty }_{i=0}\lambda ^{-2i}g(B^{i}s).\]
Thus, the graph of \( \nu  \) versus \( s \) for \( B>\lambda  \) has the
form of Weierstra\( \beta  \)'s famous example of a continuous, nowhere-differentiable
curve.

To obtain (\ref{DHatFromKM-PY}) in another way, we again consider the map (\ref{analPhiMap}),
(\ref{analNuMap}). We claim that (\ref{analPhiMap}), (\ref{analNuMap}) can
be regarded as a \emph{typical} system, and that the formulae of Sec. 2 should
apply to it. That is, in contrast to the existence of a symmetry for (\ref{analZMap})
{[}namely, \( \vec{z}\rightarrow -\vec{z} \) leaves (\ref{analZMap}) invariant{]},
(\ref{analNuMap}) has no special symmetry. The Lyapunov exponent corresponding
to (\ref{analNuMap}) is \( h_{\nu }=2\ln \lambda  \). Note that for the system
(\ref{analPhiMap}), (\ref{analNuMap}) {[}and also for the system (\ref{analPhiMap}),
(\ref{analZMap}){]} there are \emph{no} fluctuations in the finite time Lyapunov
exponents and thus the decay time for the system (\ref{analPhiMap}), (\ref{analNuMap})
is given by \( \tau _{\nu }^{-1}=h_{\nu } \). Noting that the Lyapunov exponents
for the cat map are \( \pm \ln B \) and applying (\ref{tauDecay}) to the three
-dimensional map (\ref{analPhiMap}), (\ref{analNuMap}), we immediately recover
Eq. (\ref{DHatFromKM-PY}). As discussed in Appendix B, this point of view can
also be exploited for the original ellipsoid system {[}rather than just for
the model system (\ref{analPhiMap}) and (\ref{analZMap}){]}.

Returning now to the full four-dimensional system, (\ref{analPhiMap}), (\ref{analZMap}),
and noting that \( SM \) is smooth along the direction that we eliminated when
we went from (\ref{analPhiMap}), (\ref{analZMap}) to (\ref{analPhiMap}),
(\ref{analNuMap}), we have that the dimension \( D_{S} \) of the stable manifold
of the invariant set (\( \vec{z}=0 \)) for (\ref{analPhiMap}), (\ref{analZMap})
is \( D_{S}=\hat{D}_{S}+1 \). The Lyapunov exponents for \( \vec{z} \) motion
in the four-coordinate system are \( \pm h_{z}=\pm \ln \lambda  \) and \( \pm h_{\phi }=\pm \ln B \)
for \( \vec{\phi } \) motion. In terms of the Lyapunov exponents, \( D_{S} \)
is
\begin{equation}
\label{DSubSFourD}
D_{S}=\left\{ \begin{array}{cc}
4-2\frac{h_{z}}{h_{\phi }}, & \textrm{ for }h_{z}/h_{\phi }\leq 1/2,\\
3, & \textrm{ for }h_{z}/h_{\phi }>1/2.
\end{array}\right. 
\end{equation}

Since \( h_{z}=1/\tau  \) for (\ref{analPhiMap}) and (\ref{analZMap}) we
see that, for \( h_{z}/h_{\phi }\leq 1/2 \), Eq. (\ref{DSubSFourD}) is the
same as Eq. (\ref{PresentUntilted}). Also, the system (\ref{analPhiMap}),
(\ref{analZMap}) has no finite time Lyapunov exponent fluctations, and, thus,
the information dimension and the box-counting dimensions are the same. Hence,
\( D_{S}=3 \) when \( SM \) is smooth (\( h_{z}/h_{\phi }>1/2 \)). This is
analogous to the situation \( r=1 \) and Eq. (\ref{Dimension_Uniform2D}) of
Sec. 3. {[}The lack of finite time Lyapunov exponent fluctuations for our model
system (\ref{analPhiMap}), (\ref{analZMap}) is reflected in the fact that
all periodic orbits in \( \Lambda  \) have precisely the same Lyapunov exponents,
namely \( \pm \ln \lambda  \) and \( \pm \ln B \).{]}

\subsection{Basin Boundary for a Map Modeling the Tilted Ellipsoid Billiard}

We now wish to investigate the structure of the stable manifold when we give
the ellipsoid a small tilt. Again, we adopt the approach of Sec. 4.2: we obtain
a rigorous result by utilizing a simpler map model that preserves the basic
features of the tilted ellipsoid case. Here we again use (\ref{analPhiMap})
but we now modify (\ref{analZMap}) to incorporate the main effect of a small
tilt. The effect of this modification is to destroy the invariance of \( \vec{z}=0 \).
Thinking of the first non-zero term in a power series expansion for small \( \left| \vec{z}\right|  \),
this invariance results because the first expansion term is linear in \( \vec{z} \);
i.e., the \( \vec{z} \)-independent term in the expansion is exactly zero.
When there is tilt this is not so. Thus we replace (\ref{analZMap}) by
\begin{equation}
\label{TiltMap}
\left[ \begin{array}{c}
z_{n+1}\\
v_{n+1}
\end{array}\right] =\left[ \begin{array}{cc}
\lambda  & f(\vec{\phi }_{n})\\
0 & \lambda ^{-1}
\end{array}\right] \left[ \begin{array}{c}
z_{n}\\
v_{n}
\end{array}\right] +\left[ \begin{array}{c}
f_{z}(\vec{\phi }_{n})\\
f_{v}(\vec{\phi }_{n})
\end{array}\right] .
\end{equation}
The simplest version of (\ref{TiltMap}) which still has the essential breaking
of \( \vec{z}\rightarrow -\vec{z} \) symmetry is the case where \( f=f_{v}=0 \).
Because setting \( f=f_{v}=0 \) greatly simplifies the analysis, we consider
this case in what follows (we do not expect our conclusion to change if \( f,f_{v}\neq 0 \)).
Thus, we have
\begin{equation}
\label{TiltMap2}
\left[ \begin{array}{c}
z_{n+1}\\
v_{n+1}
\end{array}\right] =\left[ \begin{array}{cc}
\lambda  & 0\\
0 & \lambda ^{-1}
\end{array}\right] \left[ \begin{array}{c}
z_{n}\\
v_{n}
\end{array}\right] +\left[ \begin{array}{c}
f_{z}(\vec{\phi }_{n})\\
0
\end{array}\right] .
\end{equation}
The problem of finding the stable manifold for the invariant set of the map,
(\ref{analPhiMap}) and (\ref{TiltMap2}), is now the same as for the previously
considered case of Eqs. (\ref{analPhiMap}) and (\ref{analNuMap}) {[}compare
the equation \( z_{n+1}=\lambda z_{n}+f_{z}(\vec{\phi }_{n}) \) with (\ref{analNuMap}){]}.
Thus, making use of this equivalence we can immediately write down the equation
for the stable manifold in the four-dimensional state space \( (z,v,\phi ,\omega ) \)
as
\begin{equation}
\label{TiltMapSM}
z=-v\sum ^{\infty }_{i=0}\lambda ^{-i}f_{z}(C^{i}\vec{\phi }),
\end{equation}
which is obtained from (\ref{ModelSM}) using the replacements \( \nu \rightarrow z/v \),
\( \lambda f\rightarrow f_{z} \), and \( \lambda ^{2}\rightarrow \lambda  \).
The rigorous results of Ref. \cite{KM-PY} again show that this is a continuous,
nowhere-differentiable surface for almost all \( \lambda  \) in \( 1<\lambda <B \),
and, furthermore, when this is so (\( \ln \lambda /\ln B=h_{z}/h_{\phi }\leq 1 \))
we have
\begin{equation}
\label{TiltMapSMDimension}
D_{S}=4-\frac{h_{z}}{h_{\phi }}.
\end{equation}
Also, \( D_{S}=3 \) when \( h_{z}/h_{\phi }>1 \). Since \( h_{z}=1/\tau  \),
this is the same as Eq. (\ref{PresentTilted}).

\section{Derivation of Dimension Formulae}

\subsection{\protect\( D_{S}\protect \) for Typical Systems}

While the dimension formulae presented in Sec. 2 and derived in \cite{HOY}
were found to be valid for the two-dimensional map of Sec. 3 and the ellipsoid
scatterer of Secs. 4-6, the derivation given in \cite{HOY} does not apply to
these systems. In the derivation of \cite{HOY}, a higher-dimensional generalization
of horseshoe-like dynamics is assumed. When this type of dynamics is responsible
for a fractal basin boundary, the boundary looks, locally, like a Cantor set
of smooth surfaces. The examples of this paper do not possess horseshoe-like
dynamics. Their dynamics produces basin boundaries which consist of a single
continuous surface; see Figs. \ref{fig: 2D Fractal}, \ref{fig: Basin Boundaries (EB)},
and \ref{fig: BB3D}. We proceed to derive dimension formulae using a more general
argument not restricted to an assumed particular type of boundary. The resulting
formulae are identical to those derived in \cite{HOY} and stated in Sec. 2.

Let \( R \) be the portion of the state space which contains all points within
\( \epsilon  \) of a nonattracting, ergodic, invariant set, \( \Lambda  \)
of an \( M \)-dimensional map, \( P \). Let \( sm \) be the portion of the
stable manifold of \( \Lambda  \) which is contained in \( R \). The points
in \( R \) are within \( \epsilon  \) of \( sm \) since \( \Lambda  \) is
a subset of \( sm \). If we sprinkle a large number, \( N_{0} \), of orbit
initial conditions in \( R \), then the number of orbits left in \( R \) after
\( n\gg 1 \) iterates is assumed to scale like (\ref{tauDecay}) \( N_{n}/N_{0}\sim e^{-n/\tau } \).
Let the map, \( P \), have Lyapunov exponents 
\[
h^{+}_{U}\geq h^{+}_{U-1}\geq \cdot \cdot \cdot \geq h^{+}_{1}>0>-h^{-}_{1}\geq \cdot \cdot \cdot \geq -h^{-}_{S-1}\geq -h^{-}_{S},\]
where \( U+S=M \).

We will use the box-counting definition of dimension
\begin{equation}
\label{Dimension0_2}
N(\epsilon )\sim \epsilon ^{-D},
\end{equation}
where \( N(\epsilon ) \) is the minimum number of \( M \)-dimensional hypercubes
(``boxes'') needed to cover \( sm \) and \( D \) is its box-counting dimension. 

We wish to develop a covering of \( sm \) using small boxes and determine how
the number of boxes in that covering scales as the size of the boxes is decreased.
We will look at how the linearized system dynamics distorts a typical small
box. This will help us determine how the Lyapunov exponents are related to the
dimension of \( sm \). Since we assume a smooth map, the dimension of the stable
manifold of the map is equal to that of \( sm \).

Cover \( sm \) with boxes which are of length \( \epsilon  \) on each of their
\( M \) sides. Call this set of boxes \( C \). The number of boxes in \( C \)
is denoted \( N_{0} \). Iterate each box forward \( n \) steps, where \( n \)
is large, but not so large that the linearized dynamics do not apply to the
boxes. Call the set of iterated boxes \( P^{n}(C) \). A typical iterated box
in \( P^{n}(C) \) is a distorted \( M \)-dimensional parallelopiped and has
dimensions 
\[
\epsilon e^{nh_{U}^{+}}\times \cdot \cdot \cdot \times \epsilon e^{nh^{+}_{1}}\times \epsilon e^{-nh_{1}^{-}}\times \cdot \cdot \cdot \times \epsilon e^{-nh_{S}^{-}}.\]
 Each parallelopiped intersects \( sm \) since its preimage (a box) did. (The
set \( P^{n}(C) \) does not, however, cover all of \( sm \) since points on
\( sm \) move closer to \( \Lambda  \) on forward iteration.)

To construct a refined covering of \( sm \) we begin by covering each parallelopiped
of \( P^{n}(C) \) with slabs of size 
\[
\stackrel{{U\textrm{ factors}}}{\overline{\epsilon \times \cdot \cdot \cdot \times \epsilon }}\times \epsilon e^{-nh_{1}^{-}}\times \cdot \cdot \cdot \times \epsilon e^{-nh_{S}^{-}}.\]
There are roughly \( \exp [n(h_{U}^{+}+\cdot \cdot \cdot +h^{+}_{1})] \) such
slabs. Only \( \sim e^{-n/\tau } \) of these slab are within \( \epsilon  \)
of \( sm \). Let \( C^{\prime } \) denote the set of \( N^{\prime }\sim N_{0}\exp [n(h_{U}^{+}+\cdot \cdot \cdot +h^{+}_{1}-1/\tau )] \)
slabs needed to cover the part of \( sm \) lying in \( P^{n}(C) \). Let us
define \( H=h_{U}^{+}+\cdot \cdot \cdot +h^{+}_{1}-1/\tau  \) so that \( N^{\prime }\sim N_{0}e^{nH} \).

Iterate each of the \( N^{\prime } \) slabs backward \( n \) steps. We now
have the set \( P^{-n}(C^{\prime }) \). It contains \( N^{\prime } \) parallelopipeds
of size 
\[
\epsilon e^{-nh_{U}^{+}}\times \cdot \cdot \cdot \times \epsilon e^{-nh_{1}^{+}}\times \stackrel{{\textrm{S factors}}}{\overline{\epsilon \times \cdot \cdot \cdot \times \epsilon }}.\]
The set \( P^{-n}(C^{\prime }) \) forms a covering of \( sm \). To calculate
the dimension (which is defined in terms of boxes) we cover the \( N^{\prime } \)
parallelopipeds in \( P^{-n}(C^{\prime }) \) with boxes which are \( \epsilon _{j}=\epsilon e^{-nh_{j+1}^{+}} \)
on each side. (We will choose the value of the index \( j \) below.) The number
of boxes needed to cover \( sm \), for boxes of size \( \epsilon _{j} \),
scales as 
\[
N^{\prime \prime }\sim N^{\prime }\frac{\epsilon e^{-nh^{+}_{j}}}{\epsilon _{j}}\times \frac{\epsilon e^{-nh^{+}_{j-1}}}{\epsilon _{j}}\times \cdot \cdot \cdot \times \frac{\epsilon e^{-nh^{+}_{1}}}{\epsilon _{j}}\times \left( \frac{1}{\epsilon _{j}}\right) ^{S}.\]
To cover the slabs with these boxes we need a factor of \( 1 \) boxes along
each direction of a slab which is shorter than the edge length of a box and
a factor of \( \epsilon e^{-nh^{+}_{k}}/\epsilon _{j} \) boxes along each direction
(here, the \( k^{\textrm{th}} \) direction) which is longer than the edge length
of a box. In terms of \( N_{0} \), 
\[
N^{\prime \prime }\sim N_{0}\exp \{n[(S+j)h_{j+1}+H-h^{+}_{1}-h^{+}_{2}\cdot \cdot \cdot -h^{+}_{j}-1/\tau ]\}.\]

To compute the dimension of \( sm \), we compare \( N(\epsilon )\equiv N_{0} \)
to \( N(\epsilon _{j})\equiv N^{\prime \prime } \) using (\ref{Dimension0_2}).
This gives \( N(\epsilon _{j})/N(\epsilon )\sim (\epsilon /\epsilon _{j})^{D}\sim \exp (nDh^{+}_{j+1}) \)
which yields the following \( j \)-dependent dimension esitmate,
\begin{equation}
\label{DofJ}
D(j)=S+j+\frac{H-h^{+}_{1}+h^{+}_{2}+\cdot \cdot \cdot h^{+}_{j}}{h^{+}_{j+1}}.
\end{equation}
Our definition of box-counting dimension, (\ref{Dimension0_2}), requires us
to find the minimum number of boxes needed to cover the set. Since we are certain
that the set is covered, (\ref{DofJ}) yields an upper bound for the dimension
for any \( j \). Thus, to find the best estimate of those given by (\ref{DofJ}),
we minimize \( D(j) \) over the index \( j \). Comparing \( D(j) \) to \( D(j+1) \)
yields the condition
\[
h^{+}_{1}+\cdot \cdot \cdot +h^{+}_{J}+h^{+}_{J}\geq H\geq h^{+}_{1}+\cdot \cdot \cdot +h^{+}_{J},\]
where \( J \) is the best choice for \( j \) (i.e., the choice giving the
minimum upper bound). The conjecture is that this minimum upper bound \( D(J) \)
from (\ref{DofJ}) is the actual dimension of \( sm \) for typical systems.
\( D(J) \) is the same as the result Eq. (\ref{Dimension_Stable}) presented
in Sec. 2 for \( D_{S} \). The derivation of the dimension formula for the
unstable manifold, \( D_{U} \), is similar to that just presented for the stable
manifold.

{[}The derivation just presented gives the \emph{information} dimension of the
stable manifold, not the box-counting dimension. We were considering the sizes
of typical boxes in the system and covered only those. This leaves boxes that
have atypical stretching rates for large \( n \) (a box containing a periodic
point, for example, will, in general, not stretch at the rates given by the
Lyapunov exponents) unaccounted for. What we have actually computed is the box-counting
dimension of most of the measure, which is the information dimension \cite{OttBook},
of the stable manifold. See \cite{OttBook} for discussion of this point.{]}

As derived, \( D(J) \) gives an upper bound on the dimension. We saw in the
ellipsoid example of Secs. 4-6 that the \( z\rightarrow -z \) symmetry of the
system lead to a special geometry for its stable manifold (Fig. \ref{fig: BB3D})
and the formula just derived did not apply (although it was an uppder bound).
It is the conjecture of \cite{HOY} that this formula gives not the upper bound,
but the exact dimension of the stable manifold for typical systems. This is
supported by our results for the tilted ellipsoid example.

\subsection{\protect\( D_{S}\protect \) for the Untilted (Atypical) Case}

The derivation above for typical systems gives the wrong dimension formula for
the stable manifold of the (atypical) untilted ellipsoid system. This typical-system
derivation consistently overcounts the boxes needed to cover the atypical stable
manifold of the untilted ellipsoid system. Sec. 6.1 showed that the stable manifold
had a ray-like structure (cf. Fig. \ref{fig: BB3D}) induced by the \( z \)-direction
reflection symmetry of the system. If we account for this structure we can modify
the typical-system derivation so that we do not overcount boxes. The modified
derivation produces the correct formula.

Recall that our untilted ellipsoid map (which we call \( P \) here) has Lyapunov
exponents \( \pm h_{z} \) and \( \pm h_{\phi } \) and decay time \( \tau  \).
We consider the case
\[
h_{\phi }>h_{z}+\tau >0\]

Let us cover the region of state space which is within \( \epsilon /2 \) of
\( \Lambda  \) with \( N_{0} \) boxes with edge lengths \( \epsilon \times \epsilon \times \epsilon \times \epsilon  \).
Call this set \( C \). The map, \( P \), is linear in \( \vec{z} \) (in particular,
the \( \vec{z} \) portion is of the form \( \vec{z}_{n+1}=DM(\vec{\phi })\vec{z}_{n} \))
so that the graph of \( SM \) in \( \vec{z} \) space for a given value of
\( \vec{\phi } \) is a straight line through \( \vec{z}=0 \), i.e. \( z=vg(\vec{\phi }) \).
We denote by \( sm \) the portion of the stable manifold which is contained
in \( C \). Since \( SM \) contains \( \Lambda  \), each box in \( C \)
intersects \( sm \).

Iterate these boxes forward \( n\gg 1 \) steps. They become a set distorted
of parallelopipeds which we call \( P^{n}(C) \). Each parallelopiped has dimensions
\[
\epsilon e^{nh_{\phi }}\times \epsilon e^{nh_{z}}\times \epsilon e^{-nhz}\times \epsilon e^{-nh_{\phi }}\]
and intersects \( sm \). We cover \( P^{n}(C) \) by \( N_{0}\exp [n(h_{z}+h_{\phi })] \)
parallelopipeds with dimensions
\[
\epsilon \times \epsilon \times \epsilon e^{-nh_{z}}\times \epsilon e^{-nh_{\phi }},\]
and discard all of these parallelopipeds which do not intersect \( sm \). There
are \( N^{\prime }\sim N_{0}\exp [n(h_{z}+h_{\phi }-1/\tau )] \) parallelopipeds
remaining which cover \( sm \). The portions of these paralleopipeds which
are in \( \epsilon >\left| z\right| >\epsilon e^{-nh_{z}} \) do not contain
\( sm \). Suppose some portion of \( sm \) did fall in \( \epsilon >\left| z\right| >\epsilon e^{-nh_{z}} \).
Since \( P \) is linear in \( \vec{z} \), this outlying portion of \( sm \)
would, upon \( n \) reverse iterations, map to the region \( \epsilon e^{nh_{z}}>\left| z\right| >\epsilon  \),
which contradicts the definition of \( sm \) given above (i.e. \( sm \) is
within \( \epsilon /2 \) of \( \Lambda  \)). Therefore we can discard the
portion of each parallelopiped which lies is \( \epsilon >\left| z\right| >\epsilon e^{-nh_{z}} \).
These \( N^{\prime } \) parallelopipeds now have dimensions

\[
\epsilon \times \epsilon e^{-nh_{z}}\times \epsilon e^{-nhz}\times \epsilon e^{-nh_{\phi }},\]
and are denoted \( C^{\prime } \). We now iterate the parallelopipeds in \( C^{\prime } \)
backward \( n \) times and call the resulting set \( P^{-n}(C^{\prime }) \).
This set contains \( N^{\prime } \) parallelopipeds with dimensions
\[
\epsilon e^{-nh_{\phi }}\times \epsilon e^{-2nh_{z}}\times \epsilon \times \epsilon .\]
We cover \( P^{-n}(C^{\prime }) \) (which covers \( sm \)) with hypercubes
which have as their edge length \( \epsilon e^{-nh_{\phi }} \). The number
of hypercubes needed is \( N^{\prime \prime }\sim N_{0}\exp [n(h_{z}+h_{\phi }-1/\tau -2h_{z}+3h_{\phi })] \).
The (information) dimension is again found by comparing the number of \( \epsilon  \)-sized
hypercubes needed to cover \( sm \) to the number of \( \epsilon e^{-nh_{\phi }} \)
sized hypercubes needed to cover \( sm \), 
\[
\frac{N^{\prime \prime }}{N_{0}}\sim \left( \frac{\epsilon e^{-nh_{z}}}{\epsilon }\right) ^{-D},\]
which yields
\[
D=4-\frac{h_{z}+1/\tau }{h_{\phi }}.\]
This expression is between three and four for \( (h_{z}+1/\tau )/h_{\phi }<1 \).

\section{Discussions and Conclusions}

In summary, this paper has tested and illustrated formulae giving the fractal
dimension of nonattracting chaotic sets in terms of their Lyapunov exponents
and decay time. We have done this using two examples, one, a noninvertible,
two-dimensional map with two expanding (positive) Lyapunov exponents, and the
other, a chaotic scattering billiard in three spatial dimensions. The first
example is particularly useful for understanding the transient measure. The
second example provides a striking illustration of the issue of atypicality,
and also has the benefit of potential physical realization (see below). Another
point is that our second example provides the first known case of a chaotic
scatterer with a fractal basin boundary that is a continuous, nowhere-differentiable
surface (such structures have been previously discussed for basins of attraction
of dissipative systems but not in chaotic scattering). Finally, in Sec. 7 we
have provided arguments that are more general than those previously given \cite{HOY}
for the dimension formulae of Sec. 2.

To conclude, we now offer some discussion on the possible experimental realization
of our billiard example. Our point here is that such billiard systems offer
a particularly nice way of experimentally realizing and studying basin boundaries.
In the past there have been very few experiments that have attempted to study
basin boundaries. The problem is the experimental difficulty of carrying out
the technique used in numerical work: namely, choosing a grid of initial conditions
and seeing where each orbit from each initial condition goes. To do this experimentally,
one would have to precisely prepare an initial condition, run the experiment
to see where the orbit with that initial condition goes, and do this for each
initial condition. The difficulties with this approach are that it is often
not possible to prepare initial conditions either on a fine enough grid or sufficiently
precisely to observe small scale basin boundary structures and that running
the experiment many times can be time consuming. (In addition, experimental
parameters may drift over the course of the multiple experimental runs.) Nevertheless,
in one case \cite{HCP} this program was successfully carried out. In that paper
an electrical circuit was used as the experimental chaotic system, and the authors
were able to experimentally demonstrate a riddled basin of attraction (a type
of basin for which final outcomes are particularly difficult to predict; see
Refs. \cite{AKYY}, \cite{OAKY}). In another work, Cusumano and Kimble \cite{C}
have devised a new experimental procedure for studying basin boundaries. This
procedure has advantages over the straightforward method of preparing initial
conditions on a grid. Using this method, Cusumano and Kimble have successfully
mapped out basin boundaries for experimental dissipative mechanical systems.
To our knowledge, these two works (and a billiard we discuss below) are so far
the only experimental investigations of fractal basin boundaries. This is in
contrast to the large number of papers that have experimentally realized fractal
structure in chaotic attractors. This disparity in the situations of attracting
and nonattracting chaotic sets seems to be largely due to the disparity in the
ease of experimental realization for the two cases.

We believe that billiard systems offer a convenient avenue for experimental
investigation of the various general types of basin boundaries. In particular,
we suggest an optical realization of chaotic scattering billiards in which the
billiard surface is made to be optically reflecting, and light rays play the
role of orbits. For the scatterer of Sec. 4-6, for example, imagine that we
have a tube with cross section as shown in Fig. \ref{fig: EllInBox}(b) and
mirrored inner walls. The length of the tube is \( L>2r_{\perp } \), where
\( r_{\perp } \) is the major radius of an ellipsoid with a mirrored surface.
Imagine that the tube is oriented vertically with its bottom, open end placed
on a red surface, and that the ellipsoid is suspended in the tube. In this configuration,
an observer looking in the top of the tube sees multiple reflections of the
red surface that is at the bottom of the tube. Since rays whose directions are
reversed retrace the same path, we can think of orbits from initial conditions
starting on the retina of the observer's eye passing through his pupil, bouncing
around in the scatterer, and then exiting either through the bottom (red) or
the top (not red). Thus, the observed boundary of the red region is precisely
the basin boundary that we have been discussing in Secs. 4-6. Note that to observe
that boundary we did not have to prepare many initial conditions and repeat
the experiment many times. Rather a single image giving a global picture of
the basin can be formed (replacing the observer's eye by a camera). We have
already applied this approach to a chaotic scatterer formed from four mirrored
spheres to experimentally study a different general type of basin boundary structure
known as a Wada boundary \cite{SOY} (a Wada boundary is a boundary separating
three or more regions such that every boundary point is a boundary point for
all regions). In the case of our ellipsoid billiard, such an experiment would
represent the first experimental realization of a fractal basin boundary with
the structure of a continuous, nowhere-differentiable surface, one of the small
number of typical possible basin boundary types \cite{McDGY}.

We thank J. A. Yorke and B.R. Hunt for discussion. This work was supported by
ONR and DOE.

\renewcommand{\theequation}{A\arabic{equation}}

\setcounter{equation}{0}

\section*{Appendix A: Slow Convergence of the Dimension Near the Transition Between Fractal and Nonfractal Behavior}

It is difficult to numerically measure the box-counting dimension of the chaotic
repellors studied in Sec. 3 and Secs. 4-6 for some range of the system parameter.
When the parameter is set near the value at which the repellor makes the transition
from a fractal to a non-fractal function the number of boxes needed to cover
the set scales as
\begin{equation}
\label{AnomalousScaling}
N(\epsilon )\sim \frac{[\ln (1/\epsilon )]^{\frac{1}{2}}}{\epsilon ^{D_{0}}}
\end{equation}
with \( \epsilon  \) (the width of a box) down to very small \( \epsilon  \).
In (\ref{AnomalousScaling}), \( D_{0}=1 \) in the case of the two-dimensional
map and \( D_{0}=3 \) in the case of the ellipsoid system. Due to the logarithmic
term in (\ref{AnomalousScaling}) accurate numerical determination of \( D_{0} \)
can be very demanding. In the following we demonstrate that (\ref{AnomalousScaling})
holds near the transition point between a fractal and a nonfractal. In particular,
we consider (\ref{2DMap-1})-(\ref{2DMapa}) {[}\( D_{0}=1 \) in Eq. (\ref{AnomalousScaling}){]}
in the case where \( \lambda _{1}=\lambda _{2}=\lambda  \). Using the same
reasoning as for the derivation of Eq. (\ref{ModelSM}) we obtain for the invariant
set the curve

\begin{equation}
\label{Lambda_function}
y(x)=-\frac{\eta }{2\pi }\sum ^{\infty }_{i=0}\lambda ^{-1}_{i}\sin (2\pi 2^{i}x).
\end{equation}
This result is also derived in Ref. \cite{McDGY}. The box-counting dimension
of the graph of \( y(x) \) versus \( x \) is \cite{McDGY} \( D_{0}=2-[\ln \lambda /\ln 2] \)
for \( \lambda \leq 2 \) and \( D_{0}=1 \) for \( \lambda \geq 2 \). Thus,
the graph of \( y(x) \) is a fractal for \( \lambda \leq 2 \) and is non-fractal
for \( \lambda \geq 2 \). We are interested in the transition region, \( \lambda \cong 2 \),
for which we wish to show that (\ref{AnomalousScaling}) applies.

To calculate the box-counting dimension of (\ref{Lambda_function}) we need
to know how the number of \( \epsilon  \)-width boxes needed to cover the set
scales with \( \epsilon  \). If we let the sum in (\ref{Lambda_function})
run from \( i=0 \) to \( i=n \), the resulting approximation to \( y(x) \)
is a smooth, finite length curve: 
\begin{equation}
\label{ysubn}
y_{n}(x)=-\frac{\eta }{2\pi }\sum ^{n}_{i=0}\lambda ^{-i}\sin (2\pi 2^{i}x).
\end{equation}
As \( n \) increases, the ratio of the height to the width of the oscillations
added by each term in the sum increases. For the purpose of estimating \( N(\epsilon ) \)
for some fixed value of \( \epsilon  \) it suffices to consider (\ref{ysubn})
with \( n \) sufficiently large. A practical size for the boxes in a covering
is one which scales as the width of the smallest oscillations of (\ref{Lambda_function}):
\( \epsilon _{n}\sim 2^{-n} \). This way we can be sure to resolve the smallest
details of the approximation curve (\ref{ysubn}) and be sure that higher order
approximations will not significantly alter our count of the number of boxes
in the covering. (Oscillations introduced by the \( n+1^{\textrm{st}} \) term,
for example, will be roughly half the width of a box of size \( \epsilon _{n}\sim 2^{-n} \),
or \( \sim 2^{-(n+1)} \) .)

The length of the curve (\ref{ysubn}) is
\[
\ell _{n}=\int ^{1}_{0}dx\sqrt{1+[y_{n}^{\prime }(x)]^{2}}\equiv \left\langle \sqrt{1+[y^{\prime }_{n}(x)]^{2}}\right\rangle ,\]
where \( \left\langle \bullet \right\rangle  \) denotes an average over \( x \).
For large \( n \) and \( \lambda \leq 2 \), \( \left| y_{n}^{\prime }(x)\right|  \)
is typically large and \( \ell _{n}\cong \left\langle \left| y^{\prime }(x)\right| \right\rangle  \). 

To determine \( \ell _{n} \) we examine
\[
y^{\prime }_{n}(x)=-\eta \sum ^{n}_{i=0}\left( \frac{2}{\lambda }\right) ^{i}\cos (2\pi 2^{i}x).\]
For \( \frac{2}{\lambda } \) sufficiently large, the largest term in this sum
dominates and \( y^{\prime }_{n}(x)\approx \left( \frac{2}{\lambda }\right) ^{n} \).
In this case, \( \ell _{n}\sim \left( \frac{2}{\lambda }\right) ^{n} \) and
\( N(\epsilon _{n})\sim \frac{\ell _{n}}{\epsilon _{n}} \). Taking \( \epsilon =2^{-n} \),
or \( n=-\ln \epsilon /\ln 2 \) we have that \( N(\epsilon )\sim \epsilon ^{-(2-\ln \lambda /\ln 2)} \).
The scaling is a power law in \( \epsilon  \). (It gives a box-counting dimension
of \( D_{0}=2-\ln \lambda /\ln 2 \).)

For \( \lambda <2 \), but \( \lambda \approx 2 \), the last term will not
dominate the sum, but the last several, up to \( n_{x} \), terms will dominate.
To estimate \( n_{x} \), set \( \left( \frac{2}{\lambda }\right) ^{n_{x}} \)
equal to a constant factor \( K>1 \) where \( K-1 \) is order one (e.g., \( K=2 \)).
This gives \( n_{x}\sim (2-\lambda )^{-1}\gg 1 \). If \( n\leq n_{x} \), then
all the coefficients \( (2/\lambda )^{i} \) in the sum \( y^{\prime }_{n}(x) \)
are of the same order, \( y^{\prime }_{n}(x)\sim \sum ^{n}_{i=0}\cos (2\pi 2^{i}x). \)
For a given value of \( x \) and different values of \( i\gg 1 \) the arguments
of the cosine terms in the sum are very different and vary rapidly with \( x \).
Thus, we regard the terms in the sum as random. In this case \( y^{\prime }_{n}\sim \sqrt{n} \),
and, thus, \( \ell _{n}\sim \sqrt{n} \). Again, we say that \( N(\epsilon _{n})\sim \frac{\ell _{n}}{\epsilon _{n}} \),
but now \( N(\epsilon _{n})\sim \sqrt{n}/\epsilon _{n}\sim \sqrt{-\ln \epsilon _{n}}/\epsilon _{n} \),
which is (\ref{AnomalousScaling}). If \( n>n_{x} \), then \( \left| y^{\prime }_{n}(x)\right| \sim n^{1/2}_{x}\left( \frac{2}{\lambda }\right) ^{n} \). 

For the two-dimensional example studied numerically in Sec. 3.3, the numerical
results are shown in Fig. \ref{fig: DvslambdaNumerical}. Due to the effect
discussed in this appendix we have not plotted data near the kink at \( \lambda _{1}=\lambda _{b} \).
The results for the ellipsoid system are discussed in Sec. 5 and shown in Fig.
\ref{fig: EBGraph}. A typical two-dimensional slice of the basin boundary for
this system is a continuous, nowhere-diffferentiable curve similar to the chaotic
saddle of the two-dimensional system. We thus have the same difficulty in measuring
the dimension near \( h_{\phi }\tau =1 \), Fig. \ref{fig: EBGraph}(a) and
\( h_{\phi }=h_{z}+\tau ^{-1} \) Fig. \ref{fig: EBGraph}(b).

\renewcommand{\theequation}{B\arabic{equation}}

\setcounter{equation}{0}

\section*{Appendix B: Dynamics of the Orientation of \( \vec{z} \)}

\begin{figure}
{\par\centering \resizebox*{0.9\columnwidth}{!}{\includegraphics{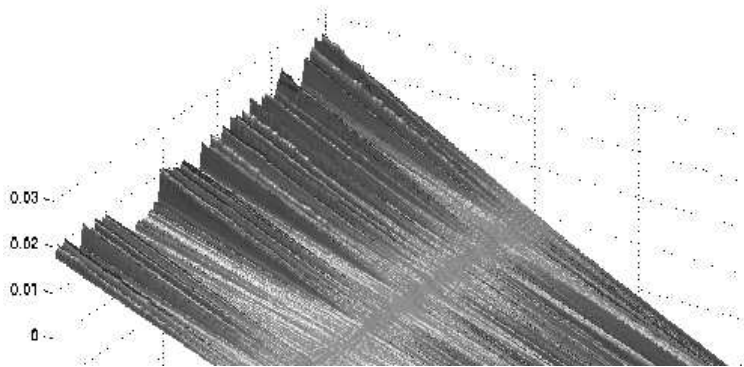}} \par}

\caption{\label{fig: BB3D}Stable mainfold (\protect\( SM\protect \)) of the chaotic
saddle (\protect\( \Lambda \protect \)); \protect\( r_{\perp }=15\protect \),
\protect\( y=5.1\protect \), \protect\( v_{x}=0\protect \), and \protect\( v_{y}\protect \)
is given by the condition \protect\( \left| \vec{v}\right| =1\protect \). The
\protect\( v_{z}=0\protect \), \protect\( z=0\protect \) line is \protect\( \Lambda \protect \).
Notice the organization of \protect\( SM\protect \) into rays emanating from
\protect\( \Lambda \protect \). This special structure is atypical and is used
to derive the atypical dimension formula for \protect\( SM\protect \), Eq.
(\ref{PresentUntilted}).}
\end{figure}

\begin{figure}
{\par\centering \resizebox*{0.9\columnwidth}{!}{\includegraphics{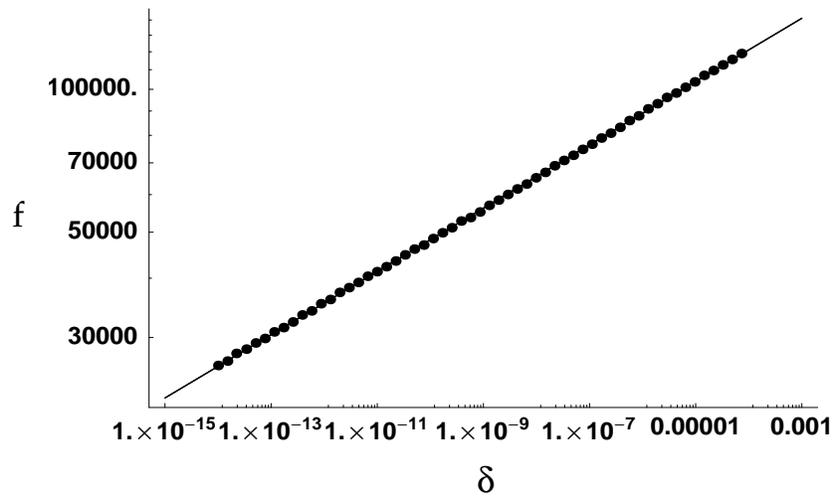}} \par}

\caption{\label{fig: fvseGraph}Uncertainty dimension measurement for the untilted ellipsoid
system with \protect\( r_{\perp }=100\protect \). Plotted is the fraction of
a random sample of points which are ``\protect\( \left| \vec{\delta }\right| \protect \)-uncertain''
for various values of \protect\( \left| \vec{\delta }\right| \protect \). The
line running through the dots is a linear fit with a slope of \protect\( \approx .07\protect \),
yielding a dimension of \protect\( d_{0}\approx 2-.07=1.93\protect \). }
\end{figure}

Rewriting (\ref{ZEqnAgain}) as,
\begin{eqnarray}
z_{n+1} & = & m_{11}(\vec{\phi }_{n})z_{n}+m_{12}(\vec{\phi }_{n})v\label{B1} \\
v_{n+1} & = & m_{21}(\vec{\phi }_{n})z_{n}+m_{22}(\vec{\phi }_{n})v_{n,}\label{B2} 
\end{eqnarray}
eliminating \( z \) in favor of \( \nu =z/v \), and dividing the first of
the above equations by the second, \( v_{n+1} \) and \( v_{n} \) cancel to
yield an evolution equation for \( \nu  \),
\begin{equation}
\label{B3}
\nu _{n+1}=\frac{m_{11}(\vec{\phi }_{n})\nu _{n}+m_{12}(\vec{\phi }_{n})}{m_{21}(\vec{\phi }_{n})\nu _{n}+m_{22}(\vec{\phi }_{n})}.
\end{equation}
Equations (\ref{PhiEqnAgain}) and (\ref{B3}) now constitute a 3D map. Our
transformation from (\ref{B1}), (\ref{B2}) eliminates the symmetry of the
original system, and it is reasonable to now assume that the information dimension
of the natural transient measure for the 3D map (\ref{PhiEqnAgain}) and (\ref{B3})
is given by the result for a typical system. Thus, we have that the information
dimension of the stable manifold of the invariant set for the 3D map is
\begin{equation}
\label{B4}
\hat{D}_{S}=3-(\hat{h}_{\phi }\hat{\tau })^{-1},\, \textrm{for}\, \hat{h}_{\phi }\hat{\tau }\geq 1.
\end{equation}
Where the superscribed circumflex denotes quantities calculated for the natural
transient measure of the 3D map, (\ref{PhiEqnAgain}) and (\ref{B3}). Since
the information dimension \( \hat{D}_{S} \) is a lower bound on the box-counting
dimension for the map (\ref{PhiEqnAgain}) and (\ref{B3}), and since the box-counting
dimension of \( SM \) for the original 4D map is one plus that for the 3D map
(this follows from the structure shown in Fig. \ref{fig: BB3D}), we have that
\( \hat{D}_{S}+1 \) {[}with \( \hat{D}_{S} \) given by (\ref{B4}){]} is a
lower bound for the box-counting dimension of \( SM \) for the 4D map. We emphasize,
however, that \( \hat{D}_{S}+1 \) is different from \( D_{S} \) for the 4D
system. This is because of the difference in the natural transient measures
for the 3D and 4D maps. (In contrast, the box-counting dimensions do not depend
on the measures.) In particular, as shown below, as compared to the 4D map natural
transient measure, the natural transient measure for the 3D map more strongly
weights \( \vec{\phi } \) values which experience slower repulsion from the
invariant set \( \Lambda  \).

To see this we consider the Lyapunov exponent corresponding to the evolution
of a differential perturbation in \( \nu  \) which we denote \( \delta \nu  \).
Differentiating (\ref{B3}) we have
\begin{equation}
\label{B5}
\delta \nu _{n+1}/\delta \nu _{n}=\bar{d}(\vec{\phi }_{n})(m_{21}\nu _{n}+m_{22})^{-2},
\end{equation}
where \( \bar{d}(\vec{\phi }_{n}) \) the determinant of (\ref{B1}), (\ref{B2}),
\( \bar{d}(\vec{\phi }_{n})=m_{11}m_{22}-m_{12}m_{21}=1 \). Noting that \( \nu _{n}=z_{n}/\nu _{n} \)
and comparing (\ref{B5}) with (\ref{B2}), we have
\begin{equation}
\label{B6}
\delta \nu _{n+1}/\delta \nu =\bar{d}(\vec{\phi }_{n})(v_{n}/v_{n+1})^{2}.
\end{equation}
For points on \( SM \), \( v_{n} \) decreases exponentially with time (i.e.,
it approaches the invariant set \( \Lambda  \)). Hence,
\begin{equation}
\label{B7}
\delta \nu _{n}/\delta \nu _{0}=Q(v_{n}/v_{0})^{-2}\sim \exp [2n\tilde{h}_{z}(\vec{\phi }_{0},n)],
\end{equation}
where \( Q=\bar{d}(\vec{\phi }_{0})\bar{d}(\vec{\phi }_{1})...\bar{d}(\vec{\phi }_{n-1}) \)
and \( \tilde{h}_{z}(\vec{\phi },n) \) with \( \vec{\phi }=\vec{\phi }_{0} \)
is the expanding \( \vec{z} \), finite-time Lyapunov exponent for the 4D system.
The estimate, \( \exp (2n\tilde{h}_{z}) \), in (\ref{B7}) follows since, due
to the Hamiltonian nature of the problem, \( Q^{1/n}\rightarrow 1 \) as \( n\rightarrow \infty  \).
Thus, for a given value of \( \vec{\phi } \), in the 3D system, the expansion
away from the invariant set is at the rate \( 2\tilde{h}_{z}(\vec{\phi },n) \)
rather than at the rate \( \tilde{h}_{z}(\vec{\phi },n) \), applying to the
4D system. For the example of (\ref{analZMap}) there are no finite time Lyapunov
exponent fluctuations {[}\( \tilde{h}_{z}(\vec{\phi },n)=\ln \lambda  \) independent
of \( n \) and \( \vec{\phi } \){]}, so that the 3D and 4D natural transient
measures are the same, both are uniform in \( \vec{\phi } \).

In this case,
\begin{equation}
\label{B8}
\hat{\tau }=\tau /2=(\ln \lambda )^{-1}.
\end{equation}
More generally, the finite time exponents, \( \tilde{h}_{z}(\vec{\phi },n) \),
fluctuate as \( \vec{\phi } \) varies. The faster expansion rate away from
\( \Lambda  \) (namely \( 2\tilde{h}_{z} \)) in the 3D case means that \( \vec{\phi } \)
values with slower expansion rates are more strongly weighted. Thus, for a case
such as our ellipsoid problem we expect that
\[
\hat{\tau }>\tau /2\, \hat{h}_{\nu }<2h_{z},\]
where \( \hat{h}_{\nu } \) is the infinite time Lyapuonv exponent for the 3D
map measure and for perturbations in \( \nu  \).

\end{document}